%% file: 3D-sensors-tb2009.tex
\documentclass[final,5p,times,twocolumn]{elsarticle}

\usepackage{graphicx}
\usepackage{amssymb}

\journal{Nuclear Instruments and Methods A}

\begin{document}

\begin{frontmatter}

\title{Test Beam Results of 3D Silicon Pixel Sensors for the ATLAS upgrade}

\author[P]{P. Grenier}
\author[X]{G. Alimonti}
\author[B]{M. Barbero}
\author[H]{R. Bates}
\author[O]{E. Bolle}
\author[M]{M. Borri}
\author[U]{M. Boscardin}
\author[H]{C. Buttar}
\author[C]{M. Capua}
\author[J]{M. Cavalli-Sforza}
\author[S]{M. Cobal}
\author[S]{A. Cristofoli}
\author[R]{G-F. Dalla Betta}
\author[G]{G. Darbo}
\author[M]{C. Da Vi{\`a}}
\author[Q]{E. Devetak}
\author[Q]{B. DeWilde}
\author[D]{B. Di Girolamo}
\author[D]{D. Dobos}
\author[L]{K. Einsweiler}
\author[S]{D. Esseni}
\author[C]{S. Fazio}
\author[T]{C. Fleta}
\author[M]{J. Freestone}
\author[D]{C. Gallrapp}
\author[L]{M. Garcia-Sciveres}
\author[G]{G. Gariano}
\author[G]{C. Gemme}
\author[S]{M-P. Giordani}
\author[O]{H. Gjersdal}
\author[K]{S. Grinstein}
\author[V]{T. Hansen}
\author[V]{T-E. Hansen}
\author[P]{P. Hansson}
\author[P]{J. Hasi}
\author[A]{K. Helle}
\author[N]{M. Hoeferkamp}
\author[B]{F. H\"ugging}
\author[P]{P. Jackson}
\author[F]{K. Jakobs}
\author[W]{J. Kalliopuska}
\author[B]{M. Karagounis}
\author[P]{C. Kenney}
\author[F]{M. K\"ohler}
\author[P]{M. Kocian}
\author[V]{A. Kok}
\author[M]{S. Kolya}
\author[J]{I. Korokolov}
\author[B]{V. Kostyukhin}
\author[B]{H. Kr\"uger}
\author[D]{A. La Rosa}
\author[M]{C. H. Lai}
\author[V]{N. Lietaer}
\author[T]{M. Lozano}
\author[C]{A. Mastroberardino}
\author[S]{A. Micelli}
\author[M]{C. Nellist}
\author[W]{A. Oja}
\author[H]{V. Oshea}
\author[J]{C. Padilla}
\author[S]{P. Palestri}
\author[I]{S. Parker}
\author[F]{U. Parzefall}
\author[M]{J. Pater}
\author[T]{G. Pellegrini}
\author[D]{H. Pernegger}
\author[U]{C. Piemonte}
\author[E]{S. Pospisil}
\author[R]{M. Povoli}
\author[D]{S. Roe}
\author[O]{O. Rohne}
\author[U]{S. Ronchin}
\author[G]{A. Rovani}
\author[G]{E. Ruscino}
\author[A]{H. Sandaker}
\author[N]{S. Seidel}
\author[S]{L. Selmi}
\author[P]{D. Silverstein}
\author[O]{K. Sj{\o}b{\ae}k}
\author[E]{T. Slavicek}
\author[O]{S. Stapnes}
\author[A]{B. Stugu}
\author[Q]{J. Stupak}
\author[P]{D. Su}
\author[C]{G. Susinno}
\author[M]{R. Thompson}
\author[B]{J-W. Tsung}
\author[Q]{D. Tsybychev}
\author[M]{S.J. Watts}
\author[B]{N. Wermes}
\author[P]{C. Young}
\author[U]{N. Zorzi}

\address[A]{Bergen University, Norway}
\address[B]{Bonn University, Germany}
\address[C]{INFN Gruppo Collegato di Cosenza and  Universit\`a della Calabria, Italy}
\address[D]{CERN, Switzerland}
\address[E]{Czech Technical University, Czech Republic}
\address[F]{The University of Freiburg, Germany}
\address[G]{INFN Sezione di Genova, Italy}
\address[H]{Glasgow University, UK}
\address[I]{The University of Hawaii, USA}
\address[J]{IFAE Barcelona, Spain}
\address[K]{ICREA/IFAE Barcelona, Spain}
\address[L]{Lawrence Berkeley National Laboratory, USA}
\address[M]{The University of Manchester, UK}
\address[X]{INFN Sezione di Milano, Italy}
\address[N]{The University of New Mexico, USA}
\address[O]{Oslo University, Norway}
\address[P]{SLAC National Accelerator Laboratory, USA}
\address[Q]{Stony Brook University, USA}
\address[R]{INFN Gruppo Collegato di Trento and DISI Universit\`a di Trento, Italy}
\address[S]{INFN Gruppo Collegato di Udine and Universit\`a di Udine, Italy}
\address[T]{Instituto de Microelectronica de Barcelona (IMB-CNM, CSIC), Barcelona, Spain}
\address[U]{FBK-irst, Trento, Italy}
\address[V]{SINTEF, Norway}
\address[W]{VTT , Finland}


\begin{abstract}

Results on beam tests of 3D silicon pixel sensors aimed at the ATLAS Insertable-B-Layer and 
High Luminosity LHC (HL-LHC)) upgrades are presented. Measurements include charge collection, 
tracking efficiency and charge sharing between pixel cells, as a function of track incident angle, 
and were performed with and without a 1.6 T magnetic field oriented as the ATLAS Inner Detector 
solenoid field. Sensors were bump bonded to the front-end chip currently used in the ATLAS 
pixel detector. Full 3D sensors, with electrodes penetrating through the entire wafer 
thickness and active edge, and double-sided 3D sensors with partially overlapping bias 
and read-out electrodes were tested and showed comparable performance.

\end{abstract}

\end{frontmatter}

\input{introduction}

\input{setup}

\input{duts}

\input{efficiency}

\input{chargesharing}

\input{conclusion}

\section{Acknowledgement}

We are very grateful to the ATLAS test beam coordinator Henric Wilkens, and to the EUDET, 
CERN SPS and North Area teams for their dedicated support and work.

\end{document}

%% file: introduction.tex
\section{Introduction}
\label{sec:introduction}

The ATLAS Collaboration will install an additional pixel layer (Insertable B-Layer - IBL) 
in the current pixel detector during the LHC shutdown currently planned for 2016 \cite{ibl}. This is to 
compensate the expected performance deterioration of the innermost layer after a few years of 
running at design luminosity (estimated integrated luminosity: 300 fb$^{-1}$). Until complete 
replacement of the entire inner 
detector for HL-LHC in 2020 or later, the IBL will have to sustain an estimated radiation 
dose, including safety factors, of $5\times10^{15} \ \rm{n_{eq}/cm^2}$, or 250 Mrad. 

3D silicon pixel technology is an attractive solution for the IBL as it provides the 
required radiation hardness. The ATLAS 3D R$\&$D Collaboration has been formed to develop 
and fabricate full 3D silicon sensors with active edge and partial 3D silicon sensors with extreme 
radiation hardness for the ATLAS experiment upgrades \cite{atlas3dcoll}.

\begin{figure}[h]
\centering
\includegraphics[scale=0.3]{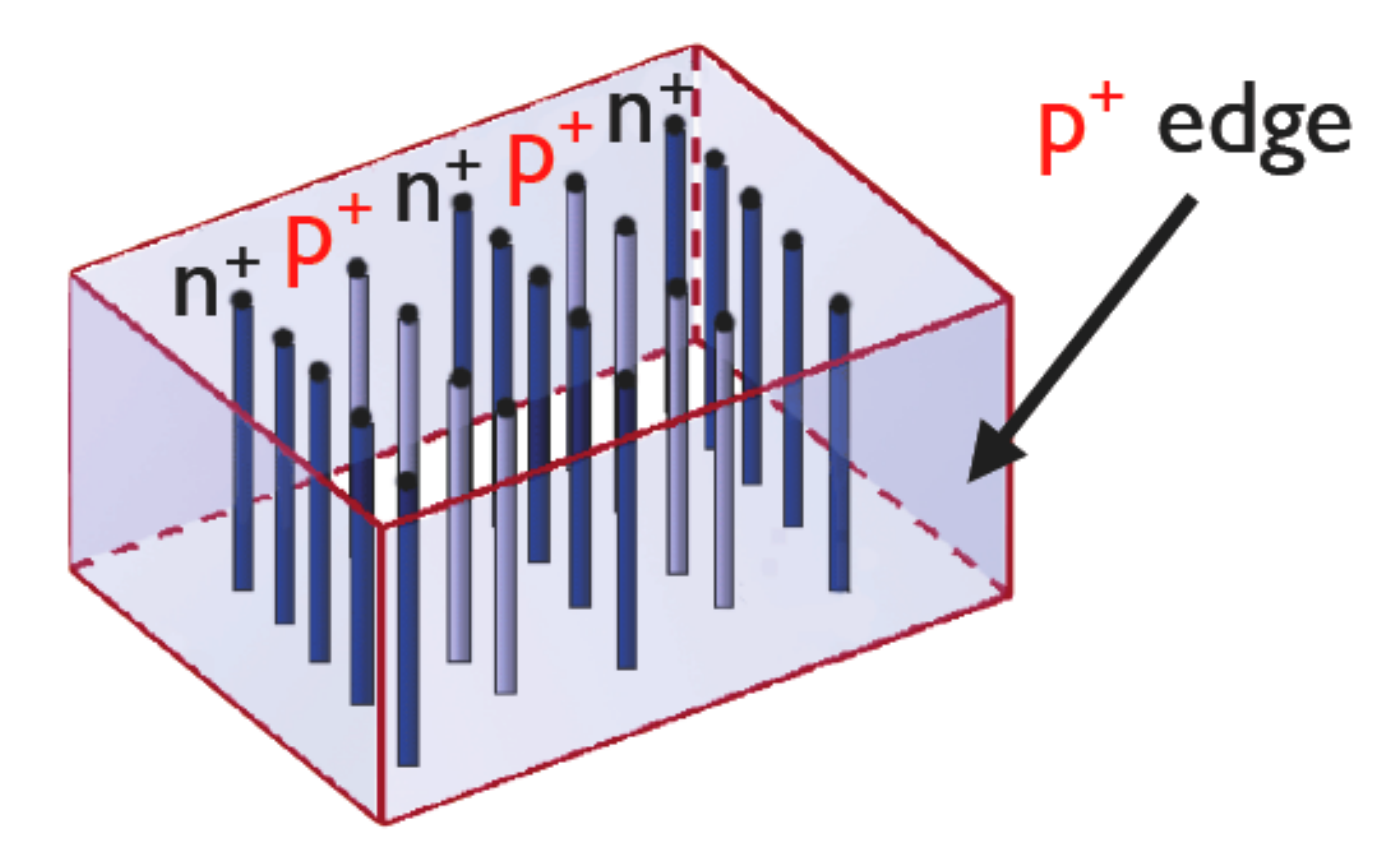}
\hfill
\includegraphics[scale=0.3]{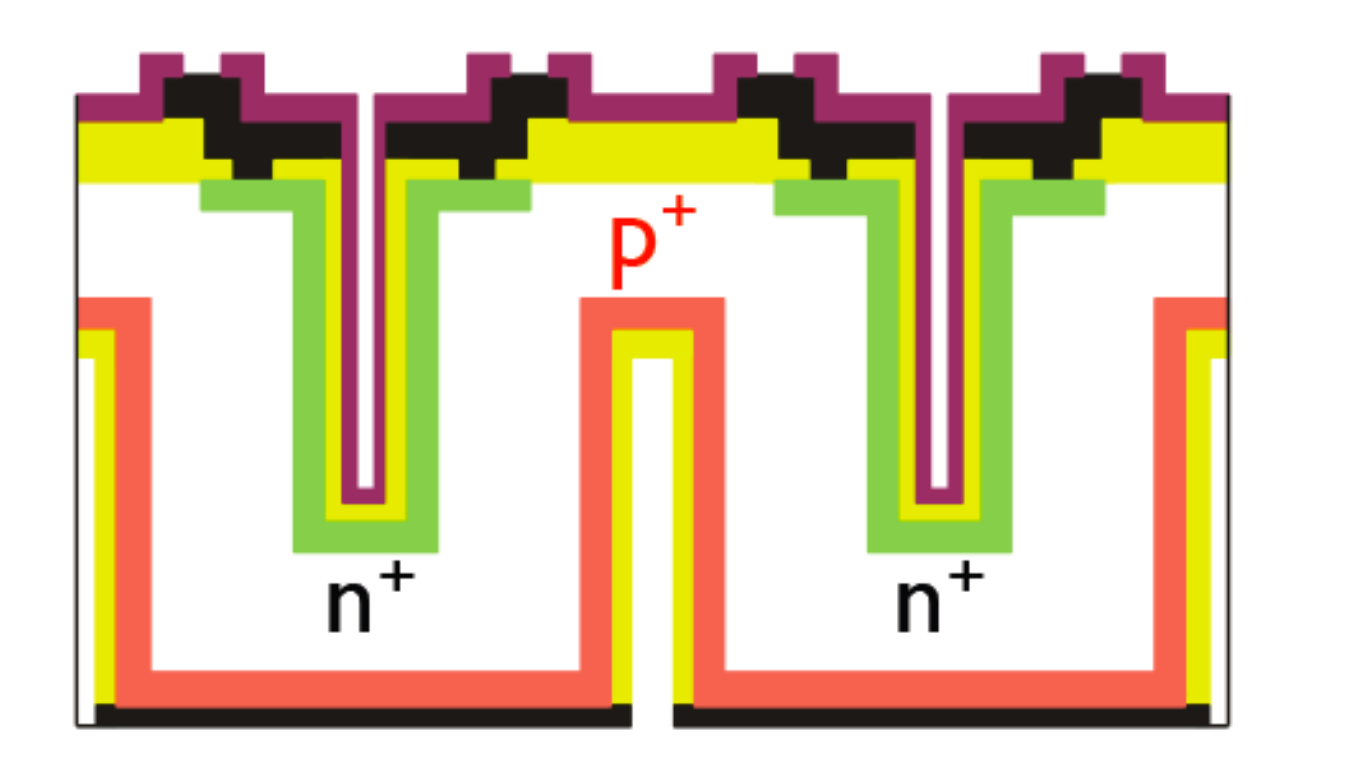}
\caption{Schematic of full-3D sensor with n$^+$ read-out and p$^+$ bias electrodes (left) 
and partial-3D (right).}
\label{two3ddesigns}
\end{figure}

Two different types of sensors are currently under evaluation: a) full-3D 
sensors with active edge and electrodes penetrating through the entire wafer thickness \cite{ref3dfull} 
and b) partial-3D where bias and read-out electrodes do not penetrate through the entire wafer thickness 
and overlap by a certain amount \cite{irstfbk} (see Figure \ref{two3ddesigns}). Both designs require 
standard VSLI processing techniques as well 
as Deep Reactive Ion Etching (DRIE) machining for electrode etching. In the case of the full-3D, 
a trench is etched, filled with p-doped polysilicon and connected to the bias electrode grid, 
making the edge of the sensor an electrode. The active edge considerably reduces the dead area 
around the sensor that is associated to planar devices.

Charge carriers drift in the silicon bulk to the read-out electrodes parallel to the wafer 
surface. The typical inter-electrode distance is 50-80 $\mu m$. The short distance between electrodes 
implies: a) fast charge collection, b) low full depletion voltage and c) short collection distance 
and consequently low charge trapping probability, and therefore high radiation hardness.

We report here beam test results on un-irradiated devices. Initial measurements on un-irradiated 
devices are necessary to assess the intrinsic performance of the sensors. Characterization of irradiated 
sensors is currently undergoing and will be reported subsequently.

%% file: setup.tex
\section{Test Beam Instrumentation}
\label{sec:setup}

Beam tests are crucial for performance characterization and optimization 
of any particle physics detector. 3D pixel sensors have been tested in beam several times 
in 2008 and 2009. Data presented here have been recorded in 2009 during 
two different periods at the CERN SPS North Area beam lines H6 (one week in October with 
a 120 $GeV/c$ $\pi^+$ beam) and H8 (two weeks in October/November with a 180 $GeV/c$ $\pi^+$ 
beam). The high momentum of the beam particles  minimizes the effect of multiple scatterring which 
is a pre-requisite for high precision tracking measurements. Previous beam tests 
results have already been reported \cite{pellenim}.

\subsection{Bonn ATLAS Telescope}
\label{subsec:bat}

\begin{figure}[h]
\centering
\includegraphics[scale=0.5]{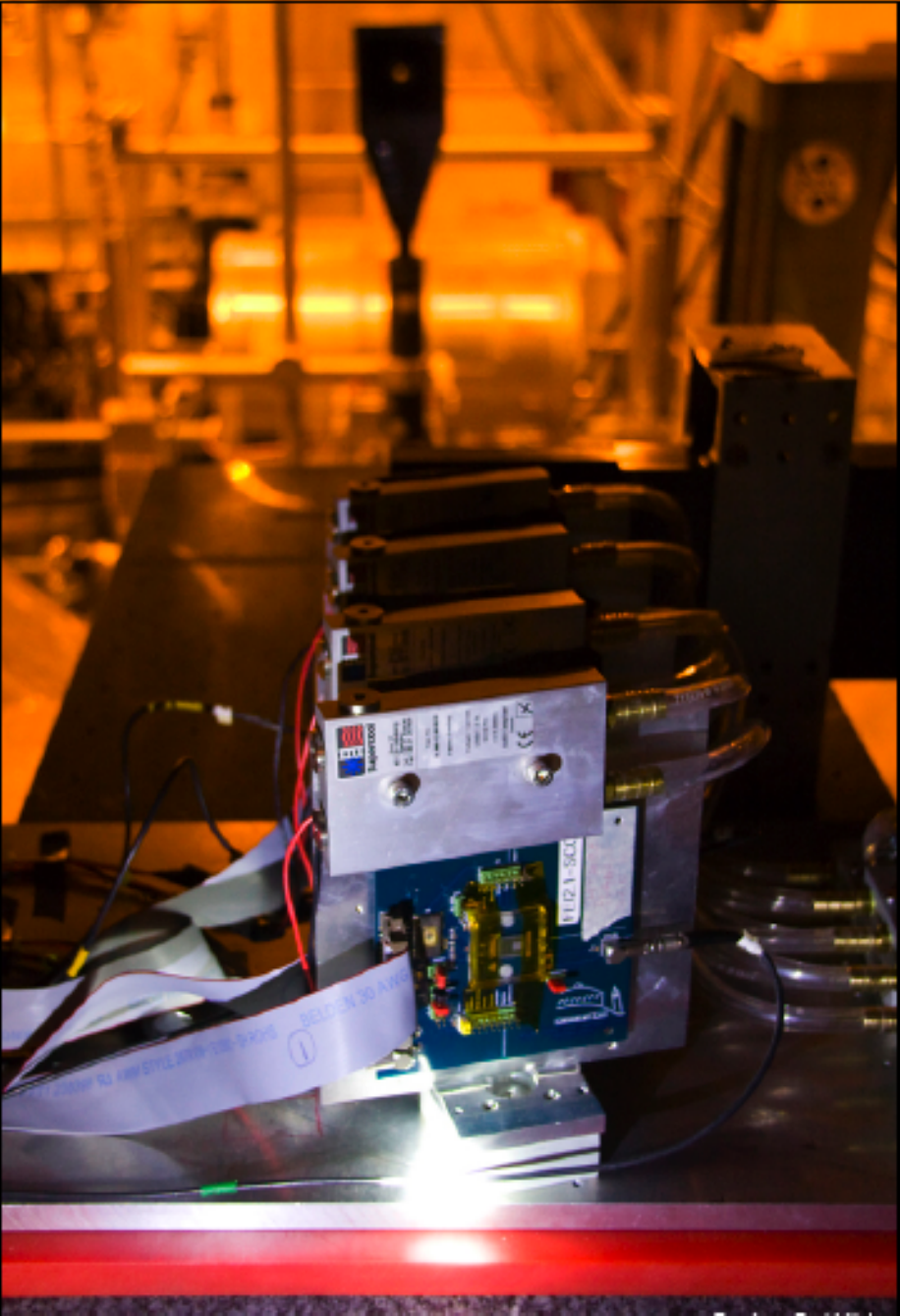}
\caption{Photo of the DUTs in the Morpurgo magnet (H8 beam line). The veto scintillator can 
be seen at the back.}
\label{photo:bat}
\vskip 0.5cm
\includegraphics[scale=0.38]{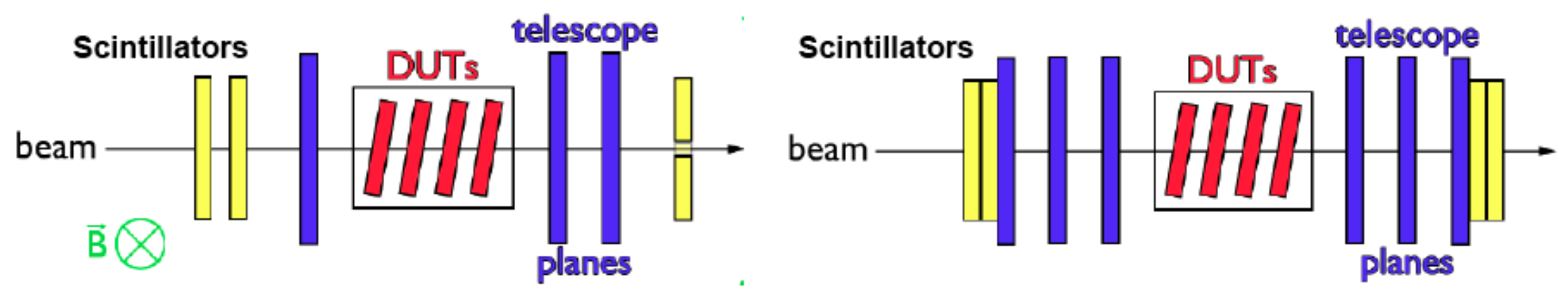}
\caption{Schematic top view of the H8/BAT (left) and H6/EUDET (right) test beam setups. Clock-wise DUTs tilt 
angles are defined as positive.}
\label{fig:bat_eudet}
\end{figure}

During the H8 data taking period the trajectories of the beam particles were reconstructed using the 
Bonn ATLAS Telescope (BAT) \cite{bat}. The telescope consisted of three $3.2 {\rm x} 3.2 \rm{cm^2}$ 
double-sided silicon strip planes with a $50 \ \mu{\rm m}$ pitch in both x and y directions. 
Tracking resolution was estimated using a full Geant4 and telescope sensor response simulation 
to be $6 \ \mu{\rm m}$ \cite{kyrrethesis}. Data acquisition 
was triggered by the coincidence of a $10 {\rm x} 10 \rm{cm^2}$ and a $2 {\rm x} 2 \rm{cm^2}$ 
scintillators located ~5 meters upstream of the Devices Under Test (DUT), and a veto scintillator 
with a ~15 mm hole located ~5 meters downstream of the DUTs (see Fig. \ref{photo:bat}). The 
purpose of the later scintillator was to suppress showering events and to reduce the data 
rate.

DUTs were mounted in the Morpurgo dipole magnet \cite{morpurgo} which provided a 1.6 T vertical 
magnetic field at the location of the sensors. Sensors were oriented with the long pixel direction 
in the vertical plane in such a way that the setup reproduced the IBL ATLAS configuration.

Data were taken at several beam incident angles, ranging from $-30^o$ to $+30^o$. DUTs tilt angle 
is described in Fig. \ref{fig:bat_eudet}. As the sensors 
were manually rotated, actual angles for each device were estimated from the alignment procedure. 

\subsection{EUDET Telescope}
\label{subsec:eudet}

In the H6 setup the high resolution EUDET Pixel Telescope \cite{eudet} was used for track 
measurement. The telescope consisted of 6 planes equally distributed into two upstream and downstream 
arms separated by about 40 cm. The core of the telescope is the Mimosa26 pixel sensor \cite{mimosa} 
with a pitch of $18.5 \ \mu{\rm m}$. Each plane consisted of 576 x 1152 pixels covering an active area of 
$10.6 {\rm x} 21.2 \ \rm{mm^2}$. Triggering was achieved by the use of upstream and downstream sets  
of two $1 {\rm x} 2 \ \rm{cm^2}$ scintillators positioned at $90^o$ with respect to each other. The 
EUDET tracking resolution has been estimated to be about $3 \ \mu \rm{m}$. 

DUTs were located  between the two telescope arms and were mounted on remotely controlled rotating 
stages. As for the H8 beam test, a tilt angle scan was performed and data were taken at several 
angles varying from $-25^o$ to $+25^o$. Larger angle values could not be reached due to 
hardware limits of the rotating stages.

A schematic of the two test beam setups is shown on Fig. \ref{fig:bat_eudet}.

\subsection{Track Reconstruction}

The two different telescopes require different approaches for track reconstruction. 

The procedure for the Bonn ATLAS telescope consists of requiring one and only one cluster in
each telescope plane. Hit positions are then estimated from these clusters, and passed to a
Kalman filter \cite{Kalman} to obtain a fitted set of track parameters. The fitted tracks are
used to align first the telescope planes, then the DUTs, using an iterative
log-likelihood method. The aligned hit estimates are then passed to the Kalman filter again,
and the track parameters are estimated at each DUT plane.

The Eudet telescope is read out in a rolling shutter mode \cite{mimosa}, meaning that one
trigger reads out hits in a time window of 112$\mu$s. This fairly large time window increases
the probability of seeing more than one track per trigger, and the amount of fake hits per
trigger due to electronic noise. In this case, making a requirement of one and only one  hit
per plane is not viable, so a combinatorial Kalman filter \cite{CKF} was implemented for track
finding and fitting. Alignment was preformed using the program Millepede II. The read-out window of the
DUTs is 400ns \cite{fei3}, so there is also a chance that a track reconstructed from the
Eudet telescope planes is out of time with the DUT read-out. In the analysis of a DUT, the
other DUTs in the beam are used to determine whether of not the track is in time with the DUT read-out.

There were some mechanical instabilities in the setup, making an accurate alignment hard to obtain for 
the full eudet data sets. For this reason data taken when the setup was most stable were selected and 
fitted with a deterministic annealing filter \cite{daf} for studies relying on good tracking resolution.

%% file: duts.tex
\section{Devices Under Test}
\label{sec:duts}

Two 3D sensors have been studied: a full-3D sensor with active edges fabricated 
at the Stanford NanoFabrication Facility \cite{stananofab} (noted as STA in the following) 
and a double-sided-double-type-column 3D sensor fabricated at IRST-FBK \cite{irstfbk} (FBK) 
with an overlap between the n$^+$ and p$^+$ electrodes of $100 \pm 20 \ \mu \rm{m}$. Wafer 
thickness was $210 \pm 10 \ \mu \rm{m}$ and $200 \pm 10 \ \mu \rm{m}$, respectively for the 
STA and FBK sensors. For the sake of reference and comparison a planar sensor (PLA) of the same 
type as the sensors currently used in the ATLAS Pixel detector \cite{atlaspixel} was also 
under test. 

All three DUTs were bump-bonded to the ATLAS Front-End Chip I3 (FE-I3) \cite{fei3}. The FE-I3 
chip is an array of 160 rows $\times$ 18 columns of $50 \mu m \times 400 \mu m$ read-out cells. 
It provides pixel charge measurement through digital time-over-threshold (TOT) measured in 
units of 25 ns, which is the LHC bunch crossing rate. Charge threshold and TOT to charge 
conversion have been tuned to 
each individual pixel to respectively 3200 e$^-$ and 60 TOT for a deposited charge of 20 ke$^-$.

\begin{figure}[h]
\centering
\includegraphics[scale=0.2]{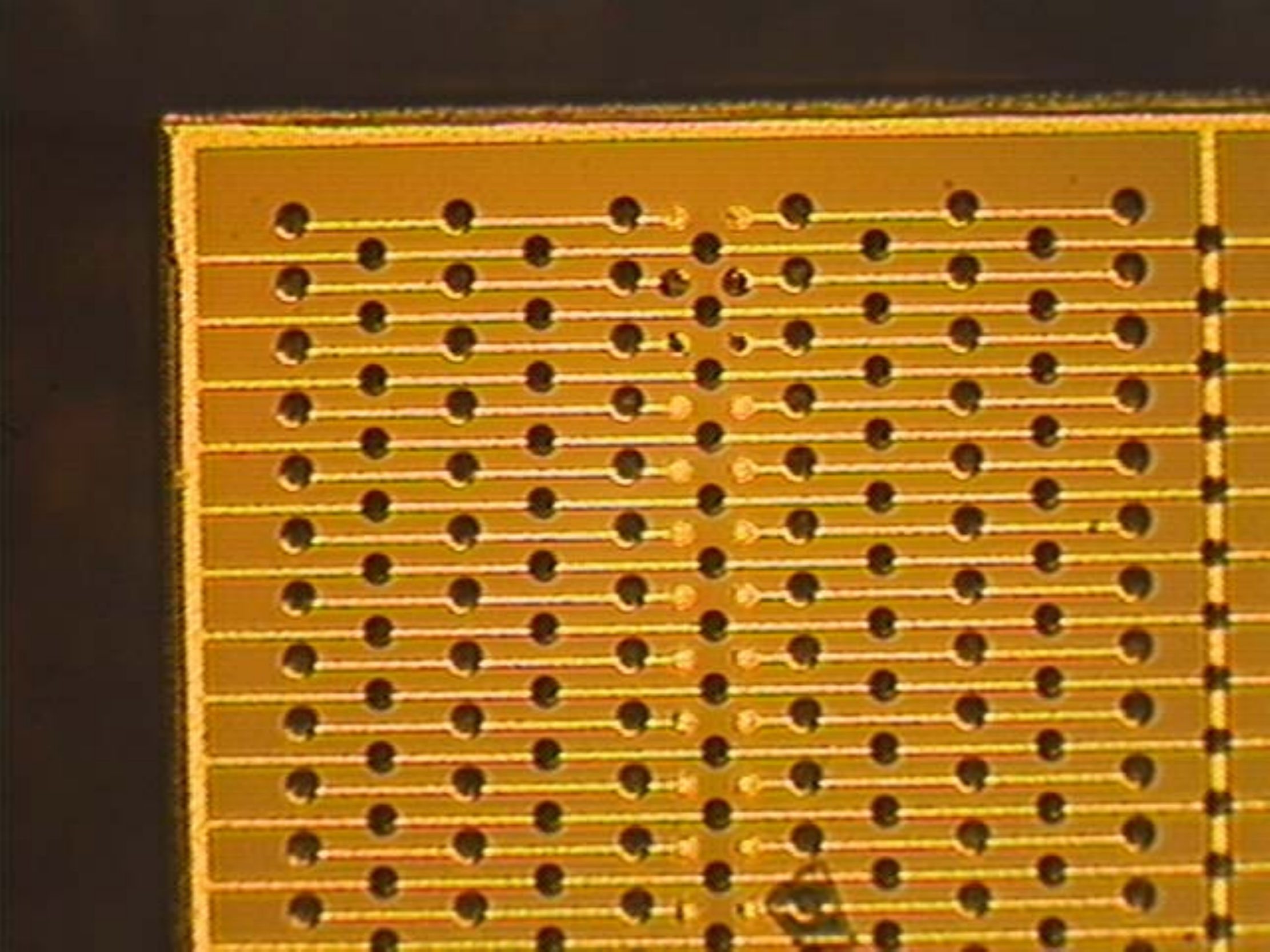}\hfill\includegraphics[scale=0.28]{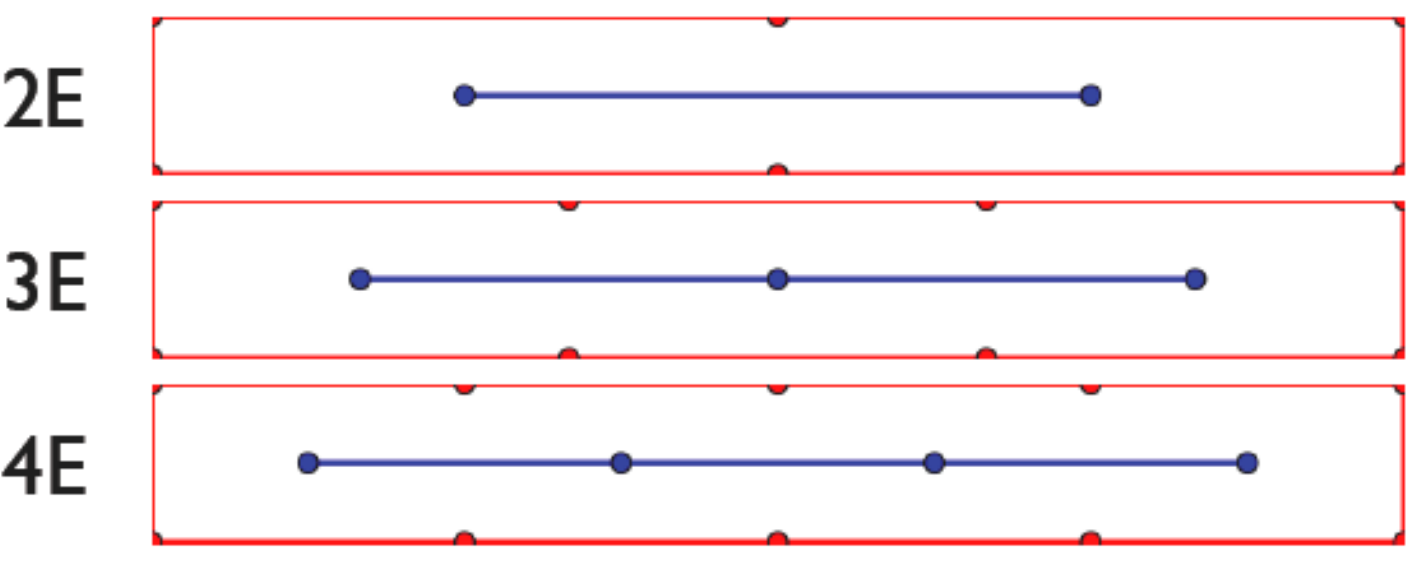}
\caption{Left: photograph of a corner of a 3D sensor with 3 electrodes per cell, showing the active edges. 
Right: 2E, 3E and 4E configurations.}
\label{fig:2e3e4e}
\end{figure}

The distance between read-out and bias electrodes and therefore the number of electrodes per 
pixel cell is an important parameter for 3D sensors as it affects key quantities such as capacitance 
and noise, bias voltage, charge collection and radiation hardness. Several configurations with 
two (2E), three (3E) and four (4E) electrodes per cell have been studied (see Fig. \ref{fig:2e3e4e}). 
The optimum configuration is 3E for an FE-I3 pixel size which corresponds to an inter-electrode 
distance of $71 \ \mu m$. The 3E configuration is considered as the best trade-off in terms of 
capacitance and charge collection efficiency at the radiation fluence expected for the IBL.

%% file: efficiency.tex
\section{Tracking Efficiency}
\label{sec:efficiency}

\subsection{Introduction}

Tracking efficiency and resolution are fundamental features of pixel detectors. Tracking 
efficiency is defined as the probability of finding a hit close to a track. Previous 
studies have shown that 3D sensors are not $100 \%$ efficient for normal incident tracks but 
recover full efficiency under a certain incident angle \cite{mathes}. This is due to signal 
loss from tracks impinging the sensor near or in the electrodes. 
 
\begin{figure}[]
\centering
\includegraphics[scale=0.4]{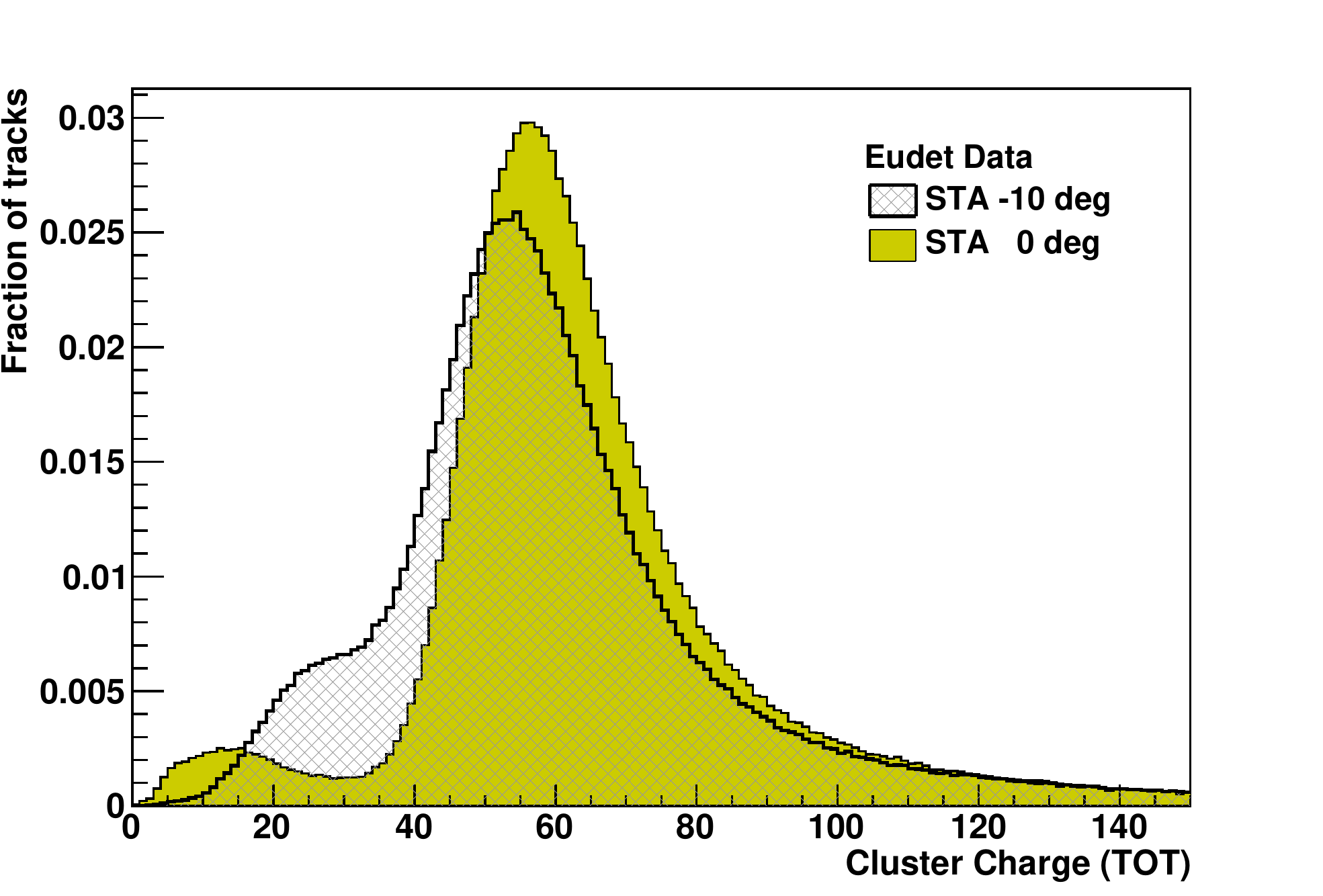}
\caption{TOT Distributions for STA sensor from the Eudet data at $0^o$ and $-10^o$.}
\label{fig:sta_tot}
\end{figure}

Following etching, thin layers of polysilicon and dopant (1-2 $\mu m$ compared to electrode 
diameter of 15-20 and 11-12 $\mu m$ for the STA and FBK sensors respectively) are deposited 
in the electrodes. In the case of the STA sensor, 
electrodes are subsequently filled with polysilicon \cite{ref3dfull}. Charge 
collection for tracks passing in the electrodes is not fully understood and it results in 
producing a lower response compared to tracks going through the bulk of the sensor. 
Given the high aspect ratio of the electrodes, track length in electrodes is small for 
inclined tracks, and enough charge is collected in the bulk region to fully recover efficiency. 
Fig. \ref{fig:sta_tot} shows the STA TOT distributions for normal and inclined ($10^o$) incident 
tracks from the Eudet data taking. At normal incidence the low TOT bump arises from tracks 
passing through the electrodes. It is attenuated for inclined tracks. 
FBK sensor electrodes are not filled with polysilicon and therefore have the aspect of empty 
holes. However, tracks are detected with high efficiency, greater than full 3D, since electrodes 
do not penetrate fully the sensor thickness. 

\begin{figure}[h]
\centering
\includegraphics[scale=0.35]{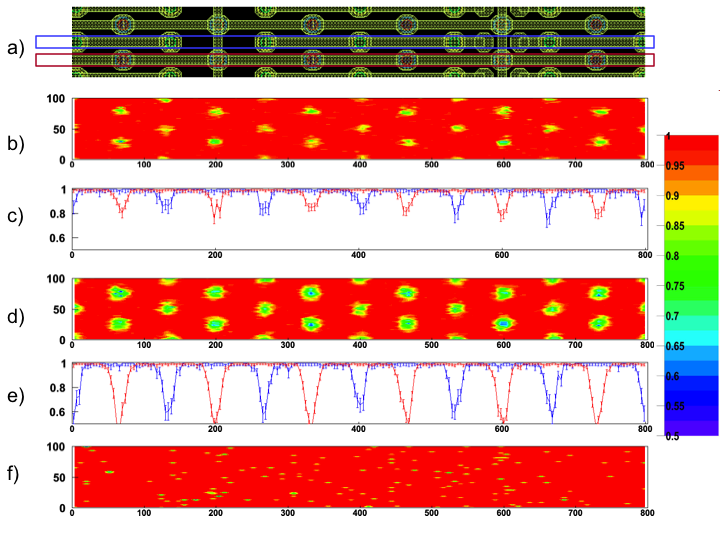}
\caption{Efficiency loss in electrodes with the Eudet data. From top to bottom: a) mask detail 
centered on one cell and extending to half a cell in both directions. b) Two-dimension efficiency map 
for the FBK sensor at normal incidence. c) FBK one-dimension efficiency projection in the read-out 
(blue curve) and bias (red curve) electrode regions, for tracks selected in the blue and red bands 
as shown in the mask, respectively. d) and e) same as b) and c) for STA. f) same 
as d) but at $-10^o$.}
\label{fig:2dmaps}
\end{figure}

Efficiency loss in the electrodes is illustrated on Fig. \ref{fig:2dmaps} which shows the 
two-dimension efficiency maps for both the STA and FBK sensors from the Eudet data at normal 
incidence and at $-10^o$ for the STA sensor. The one-dimensional projections in the read-out 
(blue) and bias (red) electrode regions are also shown. Electrode efficiencies are close to $80\%$ and 
$50\%$ for the FBK and STA sensors respectively. In the STA sensor, the tracking efficiency 
in the readout electrodes is slightly higher compared to bias electrodes. X-ray measurements 
have shown that charge collection efficiency depends on the electrode filling 
which is different for the two types of electrodes \cite{elecfill}.

\begin{figure}[h]
\centering
\includegraphics[scale=0.2]{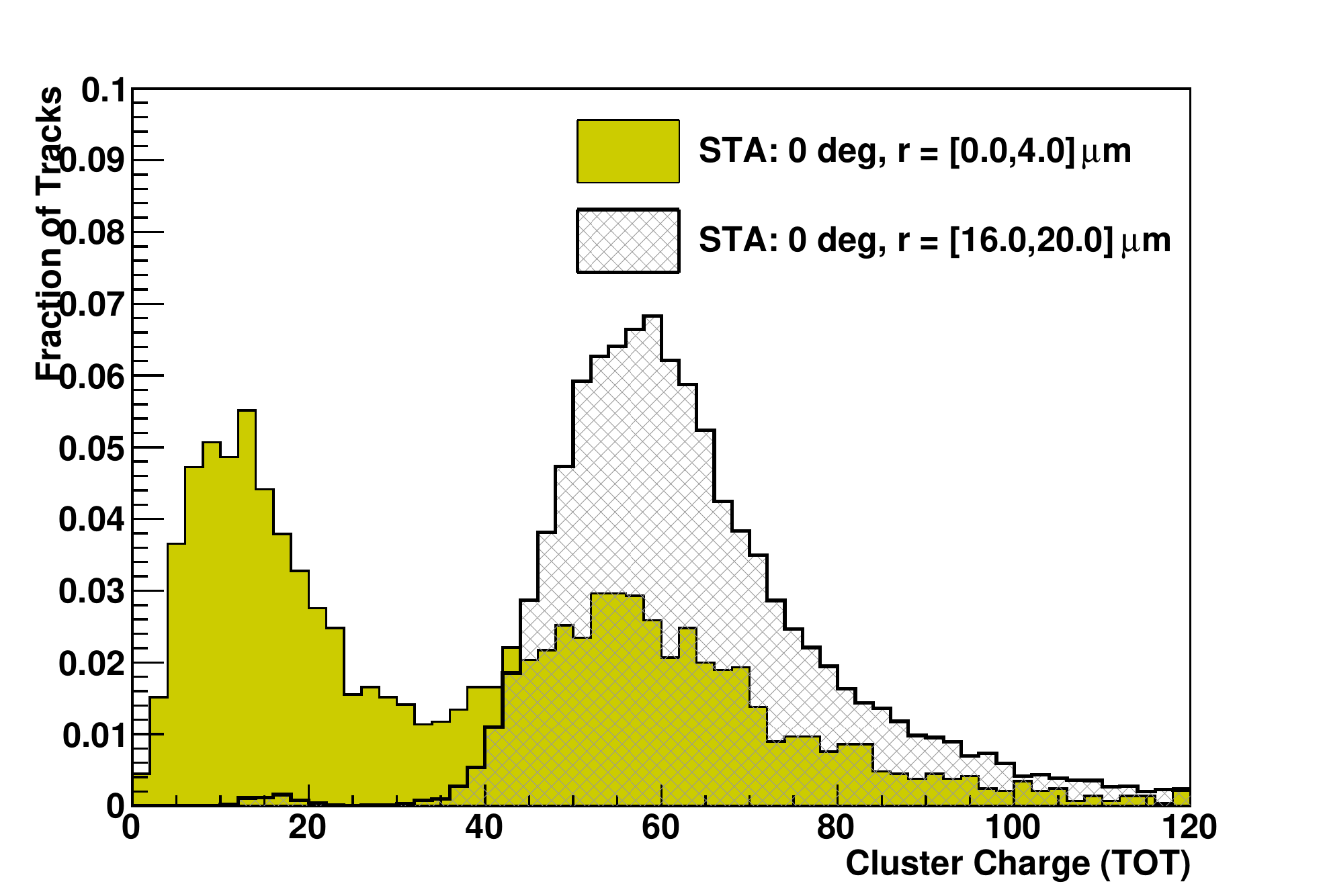}
\includegraphics[scale=0.2]{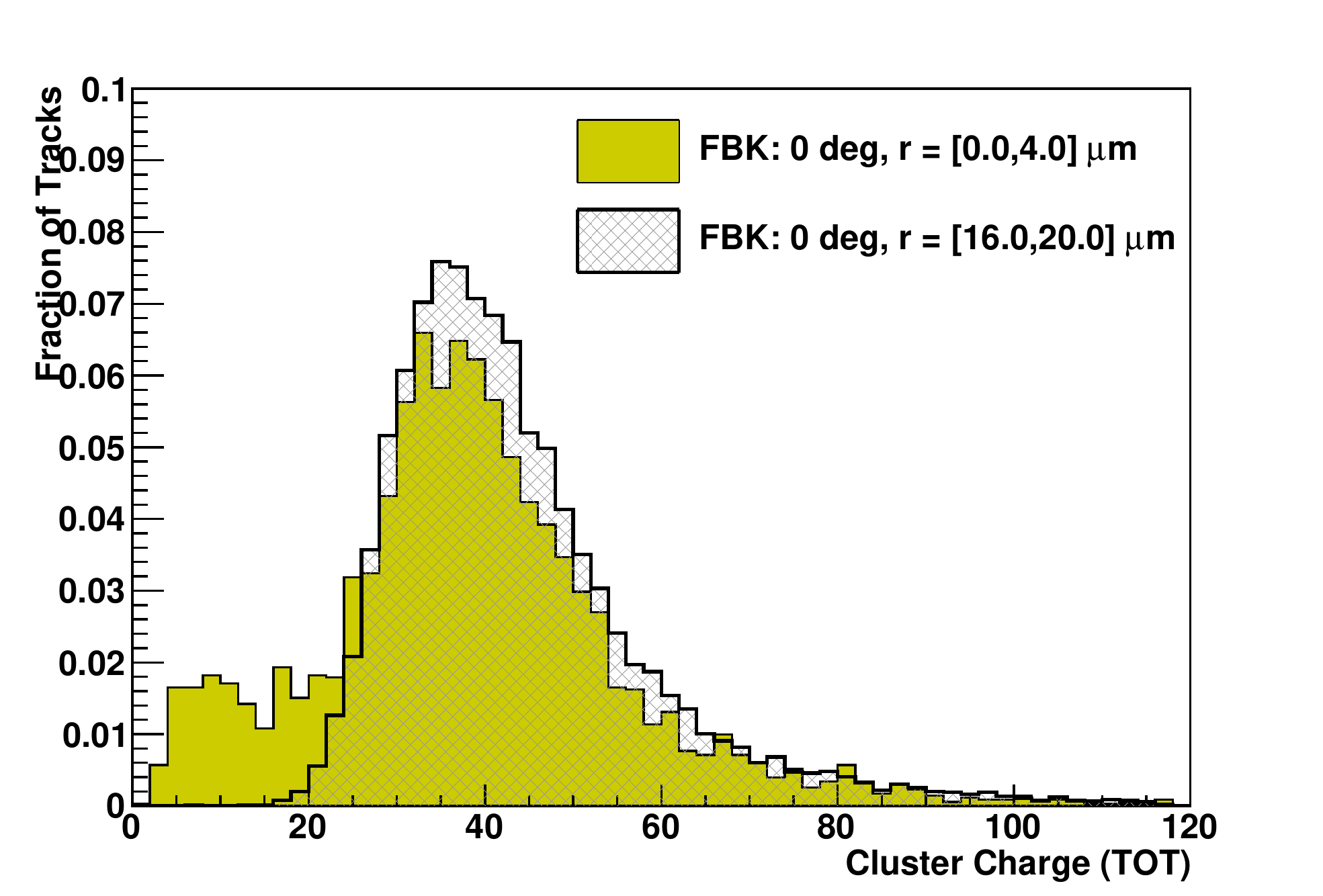}
\includegraphics[scale=0.2]{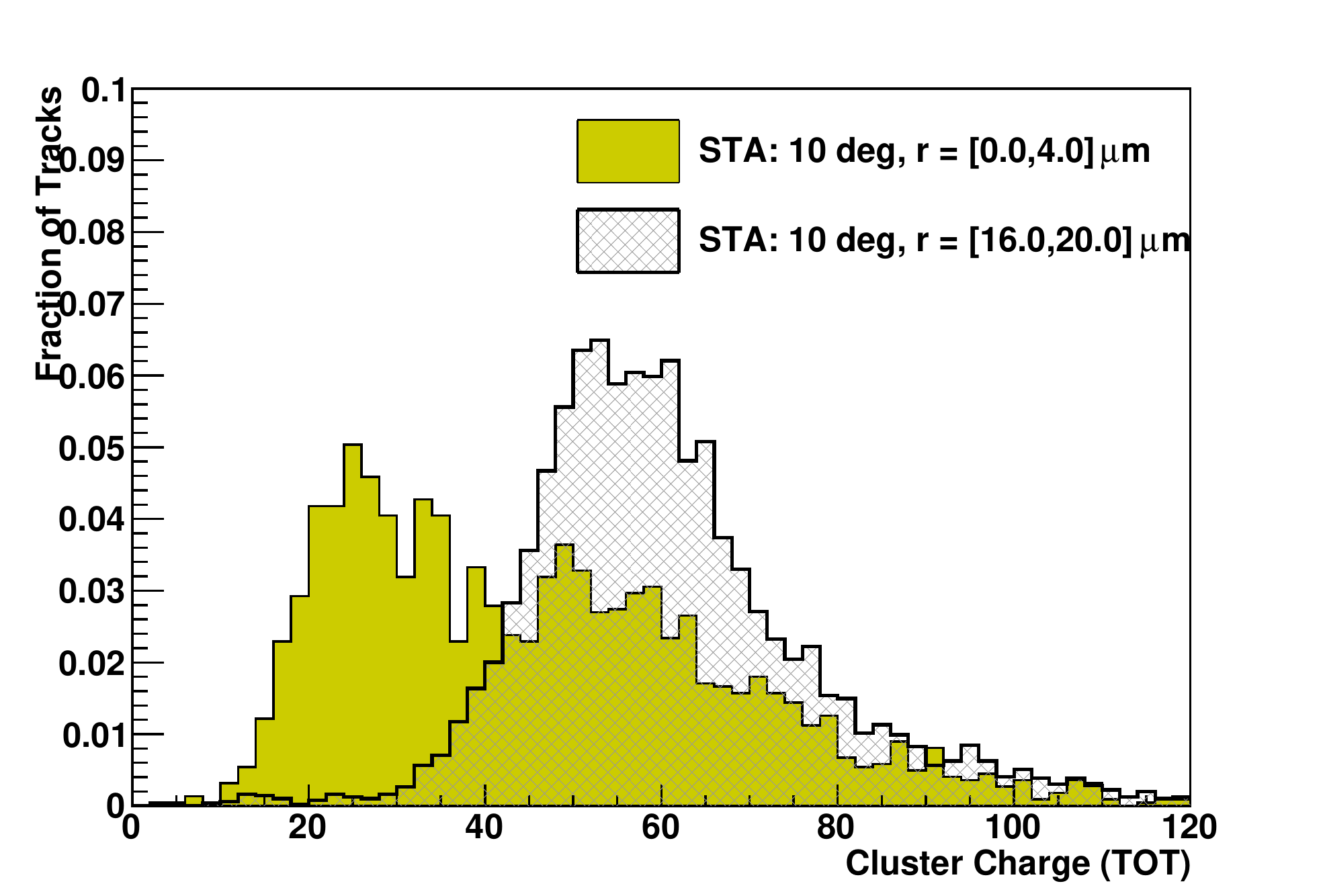}
\includegraphics[scale=0.2]{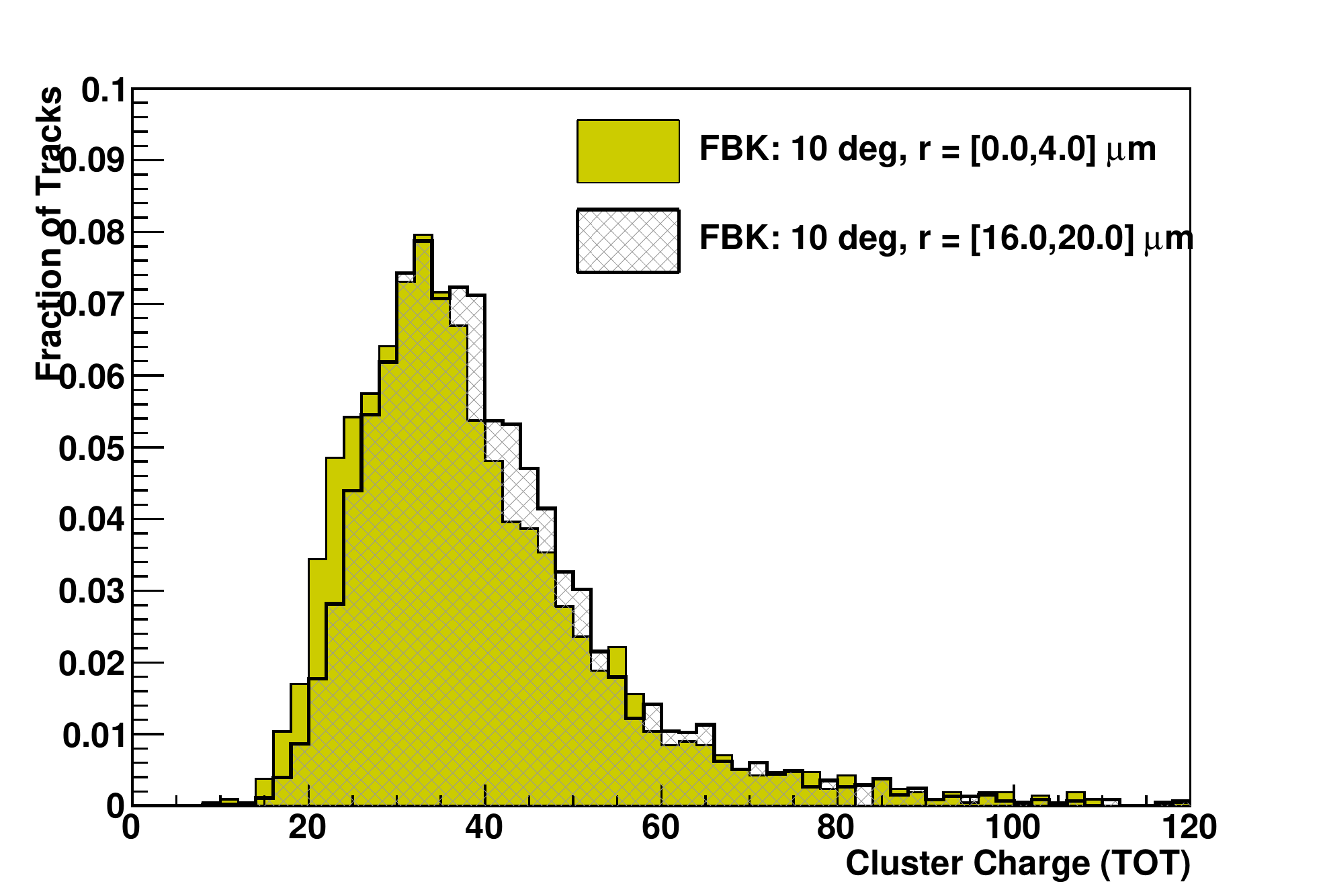}
\caption{TOT distributions for tracks going inside (yellow histograms) and outside (grey histograms) the 
electrodes for the STA and FBK sensors, at 0 and -10 degrees, from the Eudet data. Due to early breakdown 
problems, the FBK sensor was biased at a voltage not permitting full depletion of the substrate. Hence 
charge collection was not fully efficient (see Section 4.4).  }
\label{fig:XYZ}
\end{figure}

This same effect is clearly visible in Fig. \ref{fig:XYZ}: For tracks passing directly through 3D 
electrodes at $0^{\circ}$ incidence, less charge is collected, as shown by the low-ToT bump, so fewer 
tracks pass the charge threshold. The end result is an overall loss in efficiency. The FBK sensor is 
affected less strongly than the STA because its electrodes do not fully penetrate the sensor, permitting 
tracks at normal incidence to pass through at least part of the bulk. At $-10^{\circ}$ tilt, 
tracks' path length inside electrodes is significantly reduced, pushing the low-ToT bump to values 
well above threshold and fully restoring sensor efficiency. Similar performance was observed for FBK 3D 
sensors.

\subsection{Effect of magnetic field on 3D sensors}

\begin{figure}[h]
\centering
\includegraphics[scale=0.7]{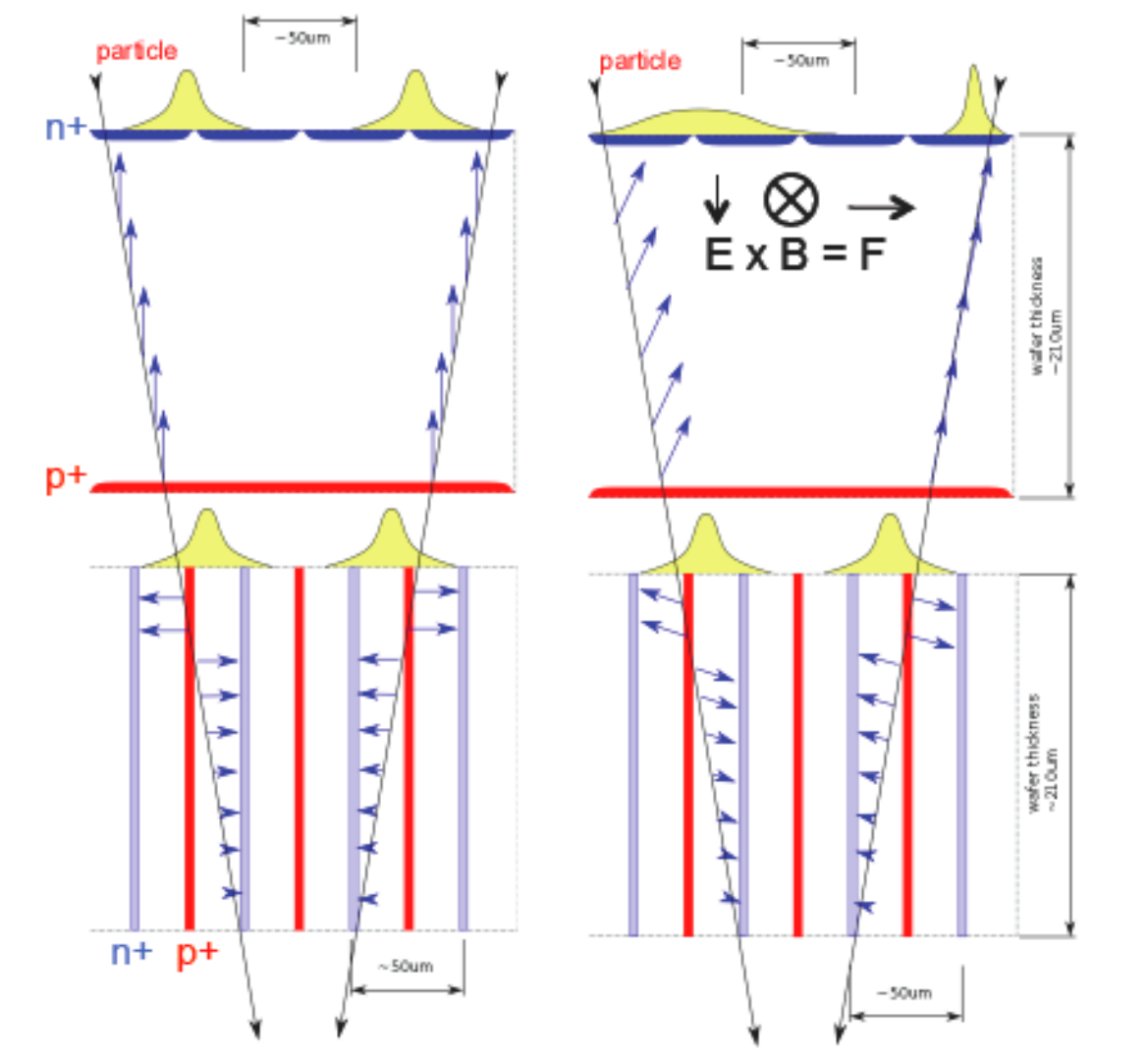}
\caption{Effect of magnetic field (left: no field and right: field ON) on planar (top) and 
3D (bottom) pixel sensors. }
\label{fig:bfield}
\end{figure}

In the ATLAS pixel detector the solenoid produces a 2T magnetic field that is orthogonal to 
the sensors' electric field. Depending on the particle incident angle the Lorentz force 
either focusses or de-focusses the drifting charges in the sensor bulk. The minimum cluster 
size is achieved at the Lorentz angle value ($(-7.6 \pm 0.6)^o$  for the current 
ATLAS pixel sensors \cite{atlaslorentz}). In 3D sensors however, the magnetic and electric 
fields are co-planar which minimizes considerably the effect of the magnetic field. Only small 
effects are expected. This was confirmed by our previous measurement \cite{pellenim}. The effect 
of magnetic field on planar and 3D sensors is illustrated on Fig. \ref{fig:bfield}.

\subsection{Efficiency as a function of tilt angle}

Overall efficiencies for the three devices under test have been determined as a function of 
the tilt angle with both the Eudet (no magnetic field) and BAT (1.6T field) telescopes. 

\begin{figure}[]
\centering
\includegraphics[scale=0.4]{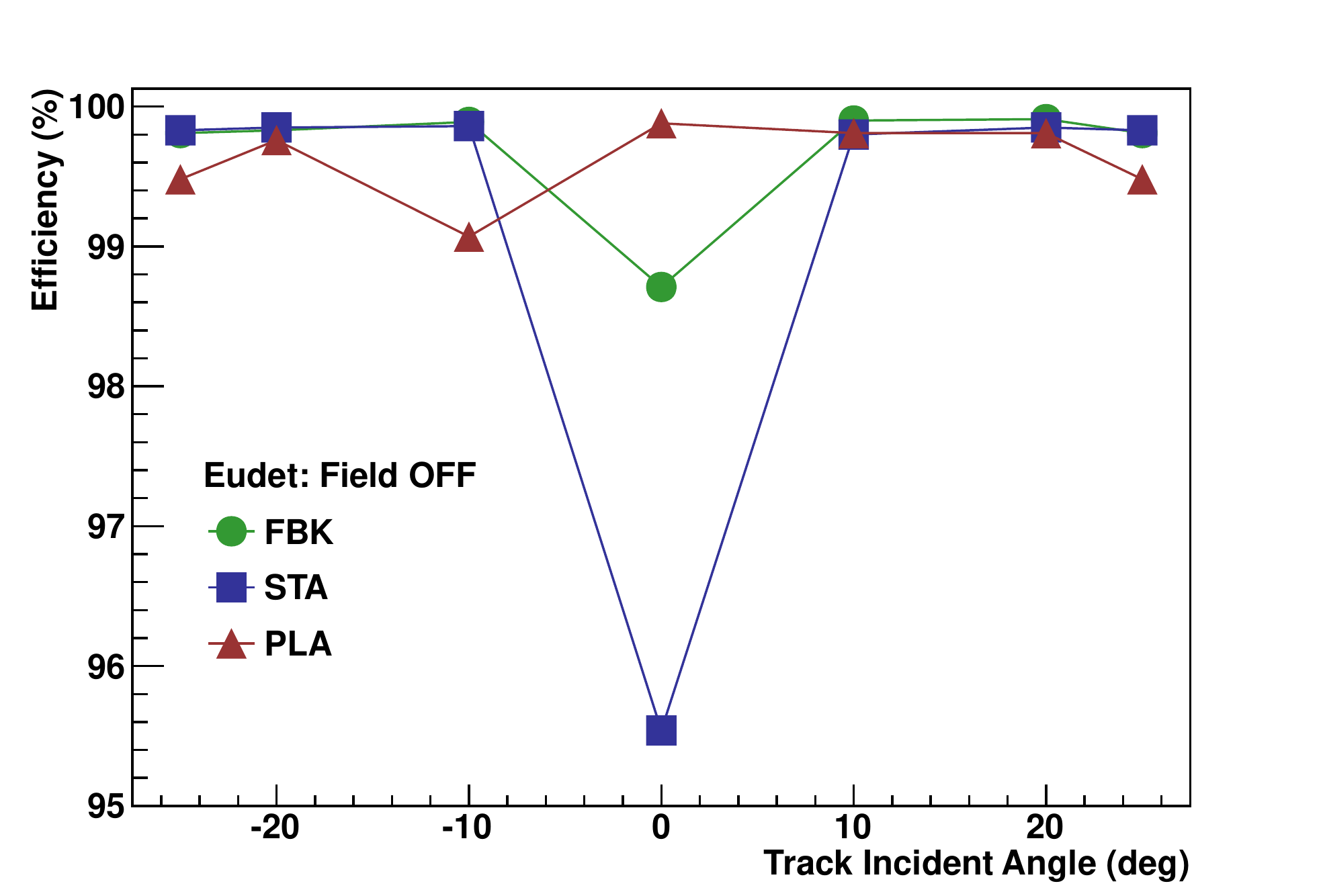}
\includegraphics[scale=0.4]{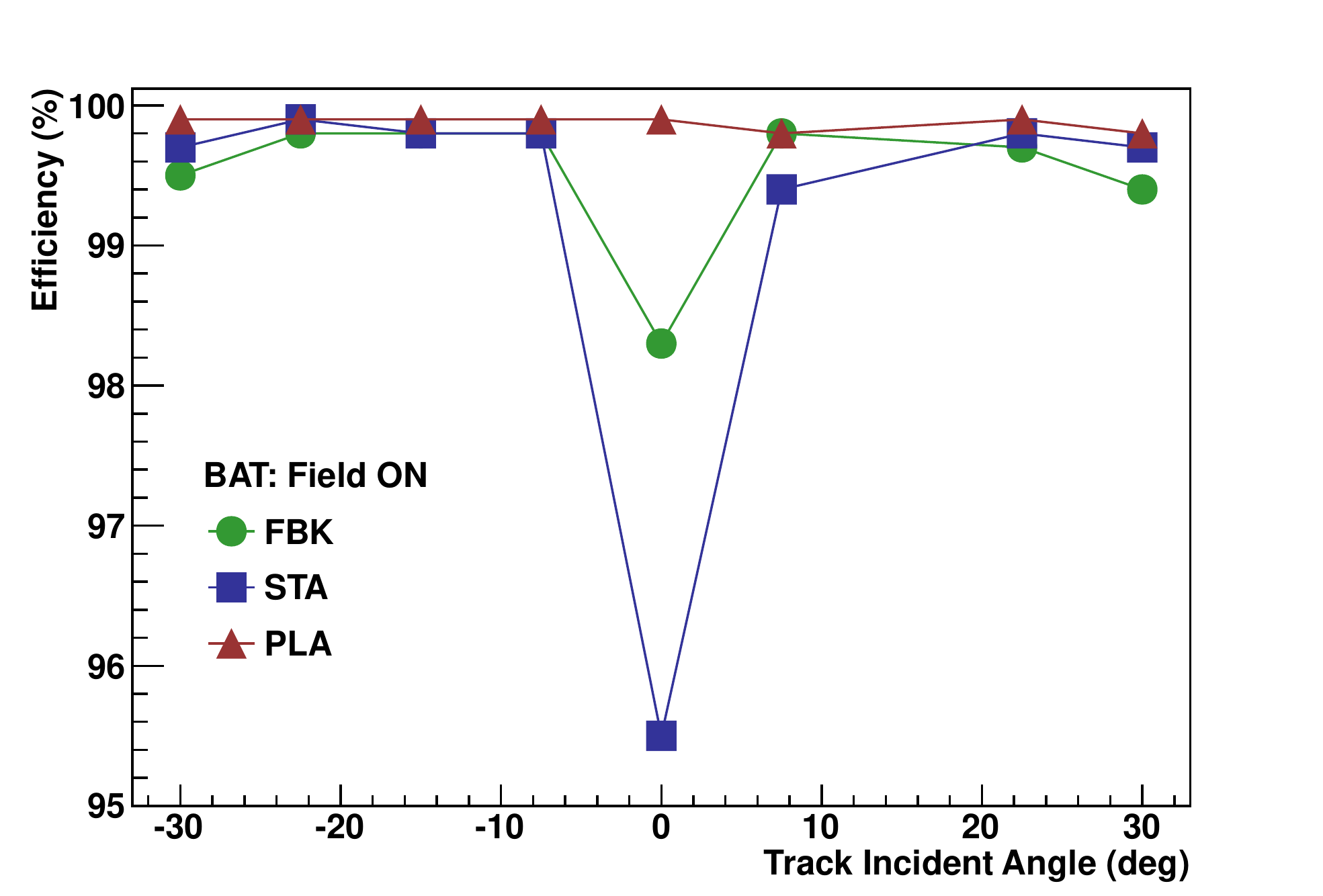}
\caption{Overall efficiency as a function of the tilt angle with field off from the Eudet data 
(top) and with field on from the BAT data (bottom). }
\label{fig:efficiency}
\end{figure}

Telescope reconstructed tracks are extrapolated to the DUTs where matching hits in space 
are checked for. In order to remove possible biases due to edge effects, only the central part 
of the sensors has been considered. Results are shown on Fig. \ref{fig:efficiency}. Systematic 
errors on the efficiencies have been estimated to $0.1\%$ from varying track selection criteria 
and track-hit matching requirements. Statistical errors are of the order of $0.1\%$ for the 
Eudet data and $0.5 \%$ for the BAT data.

\subsection{Charge collection as a function of tilt angle}

As explained above, at normal incidence the full 3D sensor has lower efficiency, 
$\epsilon = 95.5 \%$ compared to the FBK sensor $\epsilon \simeq 98.5 \%$. However the efficiency 
is fully recovered ($\epsilon > 99.5 \%$) with tilt angles greater than $10^o$. 
Consequently, both types of 3D sensors are perfectly suited to the ATLAS IBL since particle 
incident angles 
will vary from $10^o$ to $26^o$. It is confirmed that the magnetic field has very little effect 
on the performances of the 3D sensors which gave similar efficiency values for field off and on.

\begin{figure}[]
\centering
\includegraphics[scale=0.4]{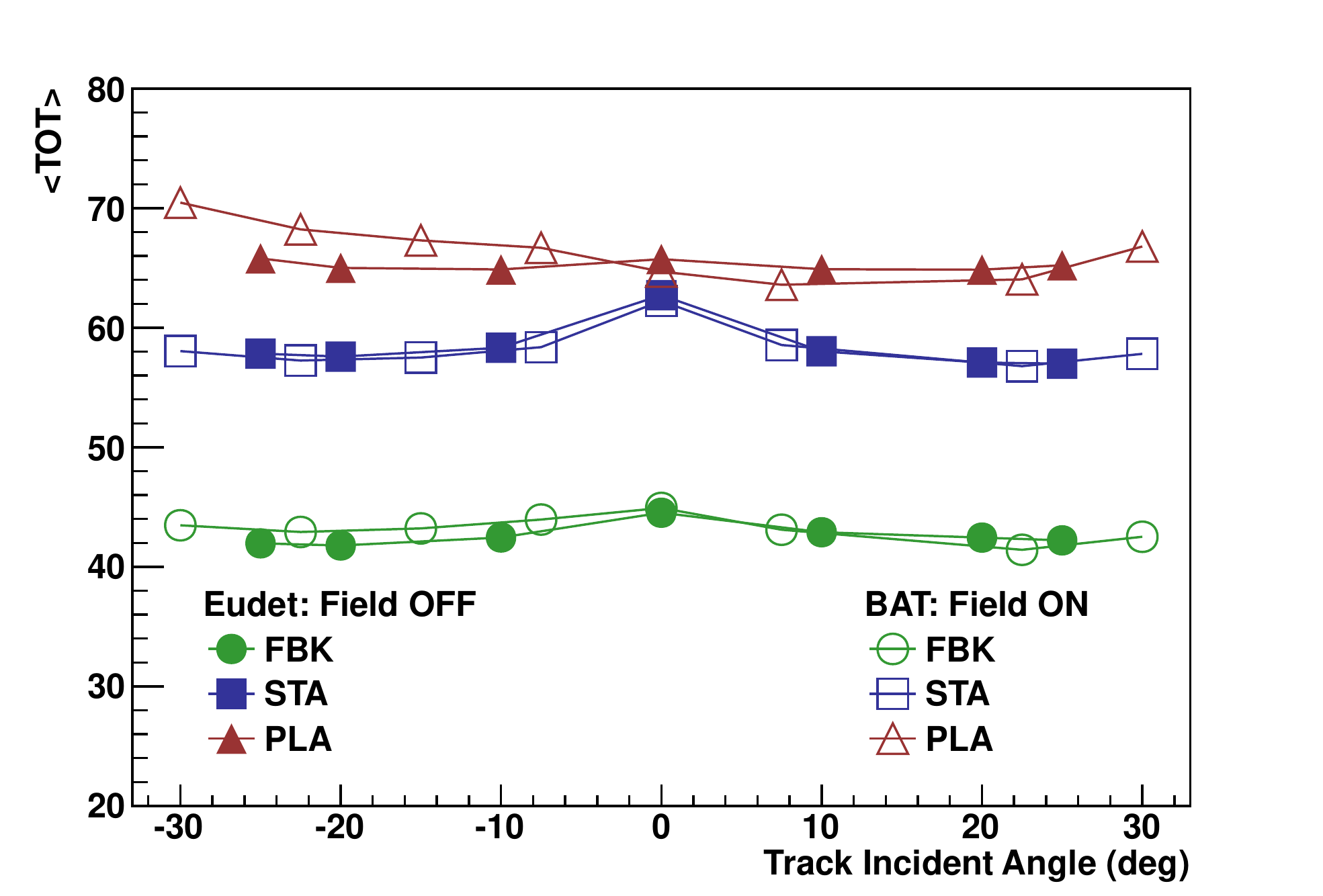}
\caption{Average of TOT distributions as a function of the beam incident angle for magnetic 
field off and on. See text for explanations of the FBK sensor lower TOT values.}
\label{fig:toth6h8}
\end{figure}

Cluster charge has been measured as a function of the beam incident angle for both setups. 
Fig. \ref{fig:toth6h8} shows the average value of the TOT distribution 
of the three sensors versus tilt angle for field off and on. At zero degree 
beam incident angle and for the two 3D sensors, charge collection is maximum as charge 
sharing is minimal (see section \ref{sec:chargesharing}). When the sensors are tilted, 
charge sharing increases and a fraction of the charge is lost in neighboring pixel cells 
that do not go over the electronics threshold. Hence, TOT decreases. At larger angles, 
the threshold effect is somewhat compensated by the longer path of particles in the silicon 
bulk which produces more charge. There is the additional effect of the Lorentz angle for 
the planar sensor in the magnetic field. A TOT increase near the Lorentz angle is visible.

\begin{figure}[h]
\centering
\includegraphics[scale=0.25]{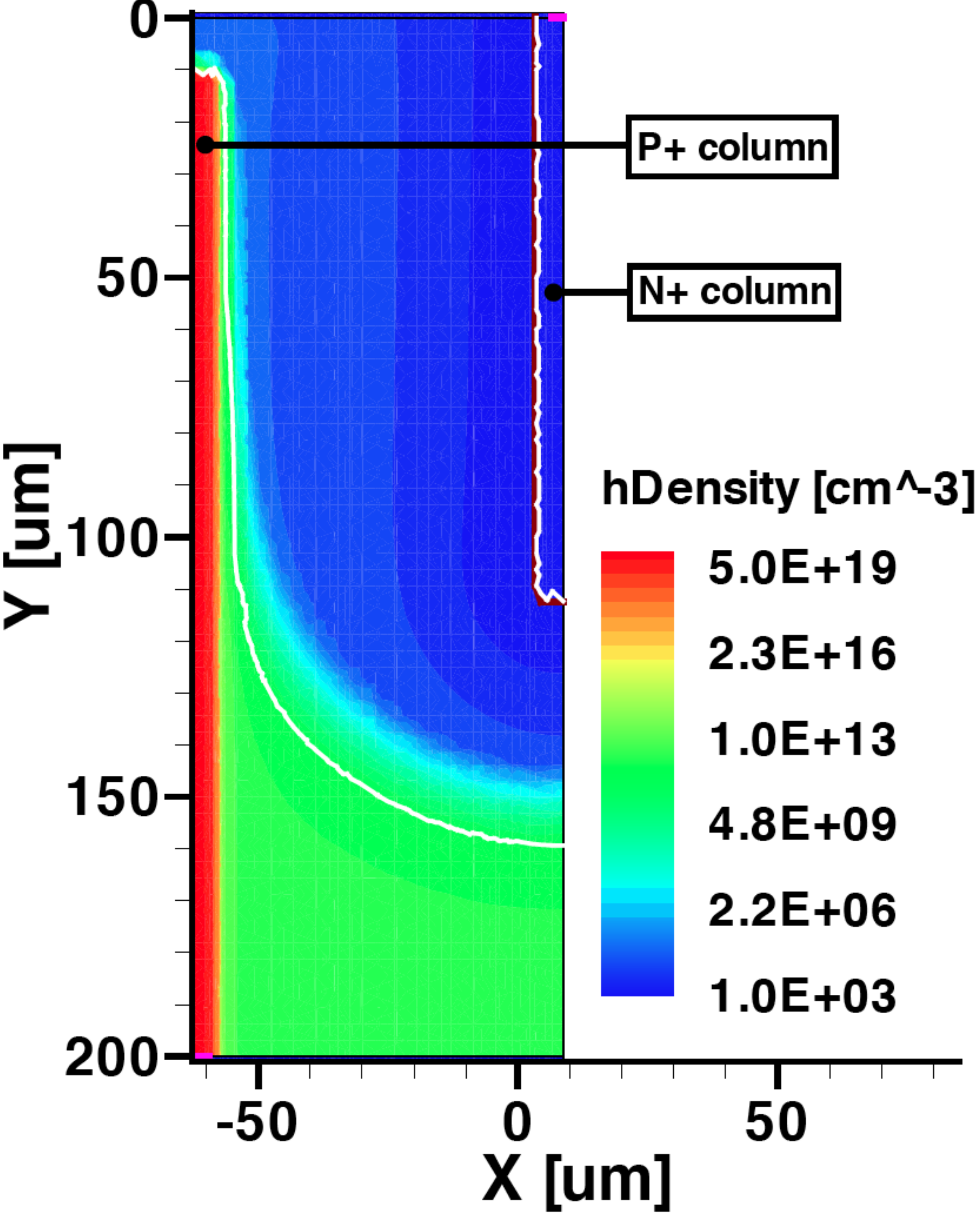}
\caption{Simulated hole density distribution along a vertical plane passing through a 
read-out (N+) column and a bias (P+) column. }
\label{gf-sim11}
\end{figure}

\begin{figure}[]
\centering
\includegraphics[scale=0.4]{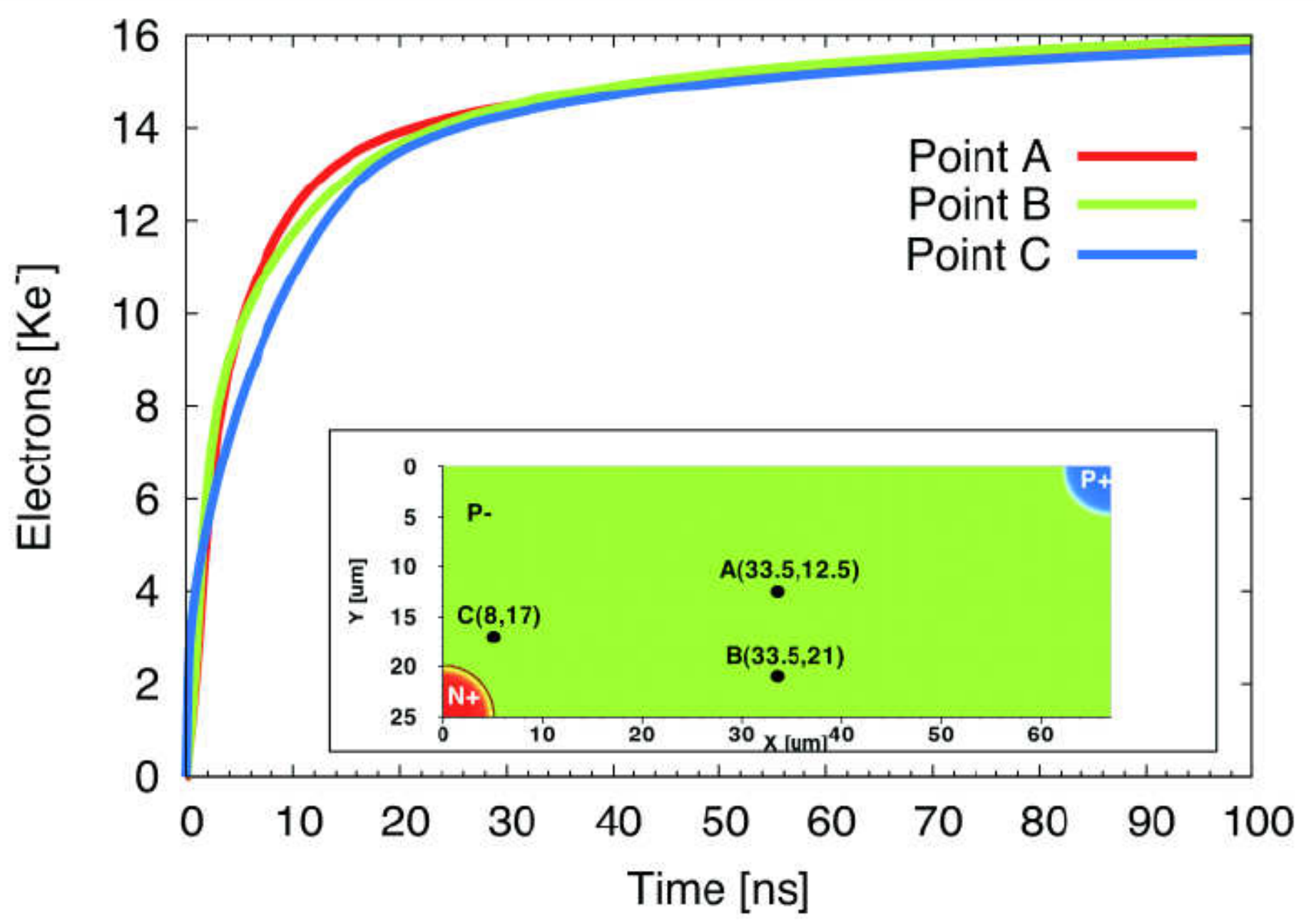}
\caption{Time integral of the simulated current pulses induced by minimum ionizing particles 
hitting the detector perpendicularly to the surface in three points. }
\label{gf-sim12}
\end{figure}

As for the FBK 3D sensor (green line with circles in Fig. \ref{fig:toth6h8}), it should be mentioned that, 
due to early breakdown problems occurring at about 10 V \cite{irstfbk}, during the beam test it was 
biased at 8V, a voltage for which the substrate is not fully depleted. This could be confirmed 
by TCAD simulations. Fig. \ref{gf-sim11} shows the simulated hole density distribution along a vertical 
plane passing through a read-out (n+) column and a bias (p+) column: as can be seen, the region 
between the two electrodes is indeed depleted, but a large portion of the substrate at the bottom 
of the device is not depleted. As a result, charge collection is expected to be rather 
inefficient from the non depleted region. This is confirmed by the data in Fig. \ref{gf-sim12}, 
showing the time integral of the simulated current pulses induced by minimum ionizing particles 
hitting the detector perpendicularly to the surface in three points shown in the inset, 
chosen as representative of different electric field conditions. The charge collected 
in 20 ns (peaking time of FE-I3 read-out circuit) is in the range from 13000 to 14000 
electrons, in good agreement with the values indicated in Fig.10.

The STA sensor was biased at 35V and was fully depleted.

%% file: chargesharing.tex
\section{Charge Sharing}
\label{sec:chargesharing}

\begin{figure}
\centering
\includegraphics[scale=0.4]{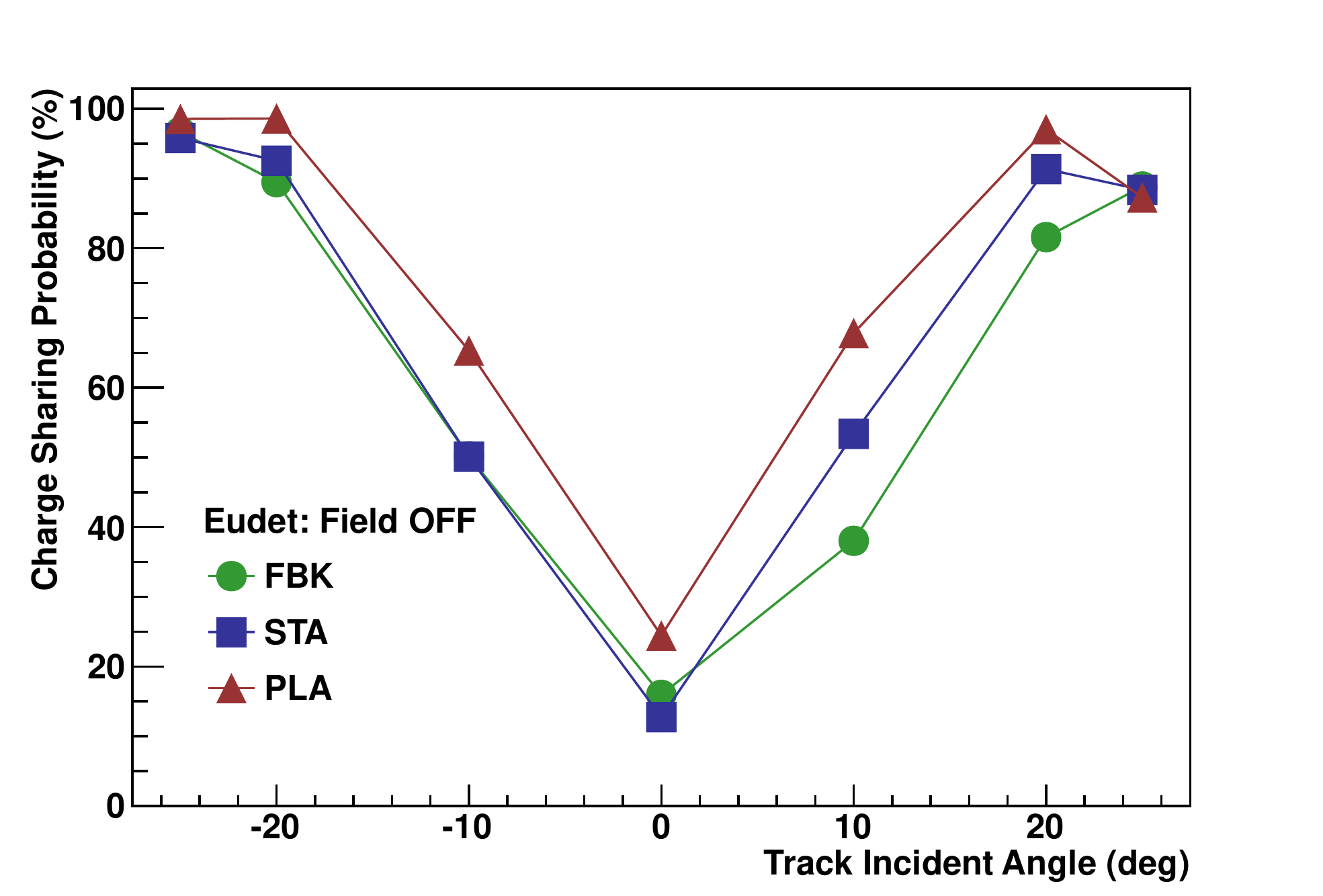}
\includegraphics[scale=0.4]{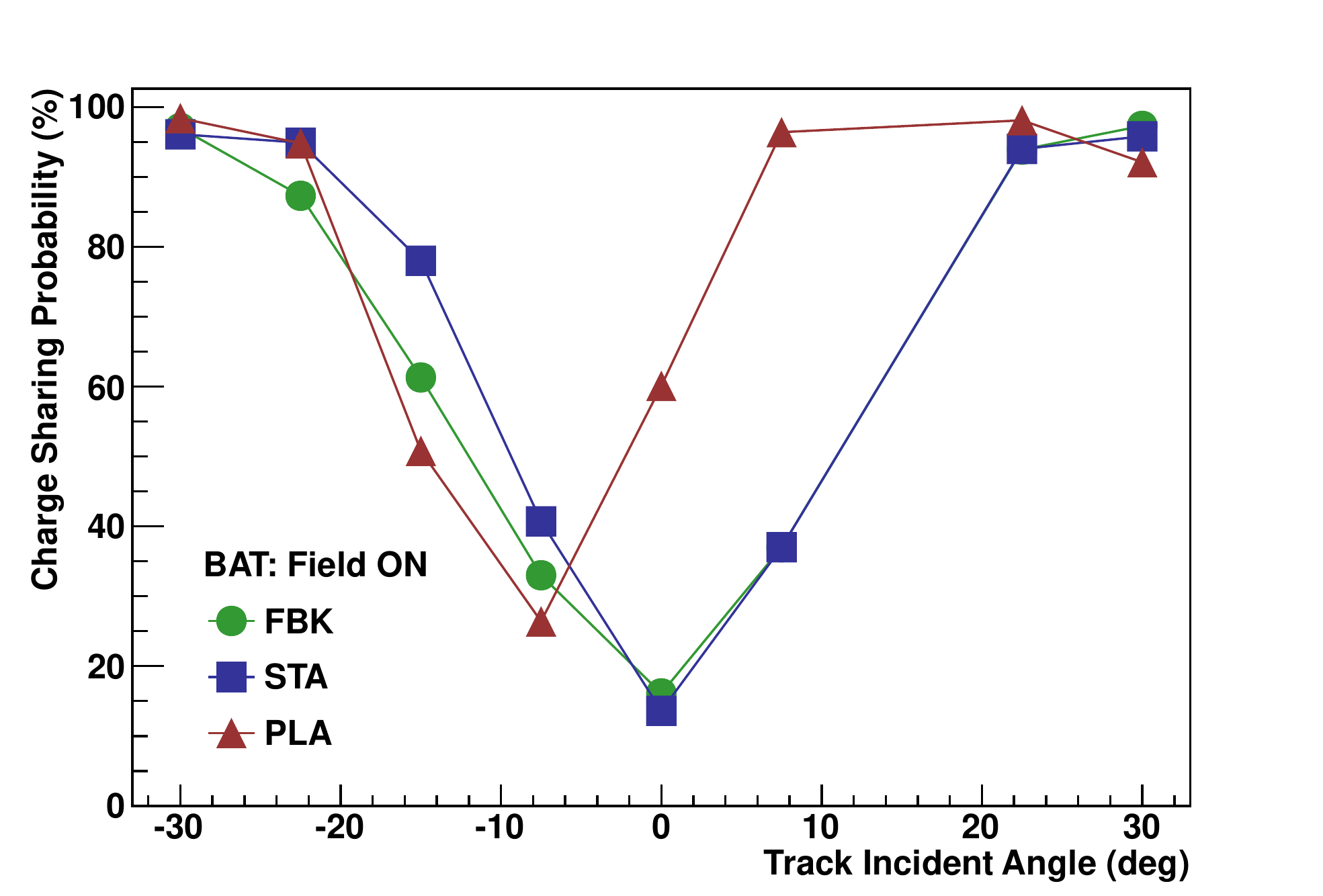}
\caption{Overall charge sharing probability as a function of beam incident angle with magnetic 
field off (top) and on (bottom). }
\label{fig:chargesharingvstilt}
\end{figure}

\subsection{Introduction}
Charge sharing is another important feature of pixel detector as it is directly 
related to tracking 
resolution and radiation hardness. The generated signal of a track going through a 
sensor can be shared between two or more cells. High charge sharing results in 
better tracking resolution as the track position can be more precisely determined. 
On the other hand, less signal will be available to each of the hit pixel cells, 
decreasing the probability to go above the comparator threshold and therefore being 
registered.

It is well known that charge collection efficiency decreases under radiation exposure. 
Hence it is desirable to minimize charge sharing for detectors running in a high radiation 
environment, such as ATLAS, in order to maintain high efficiency.

\begin{figure}
\centering
\includegraphics[scale=0.36]{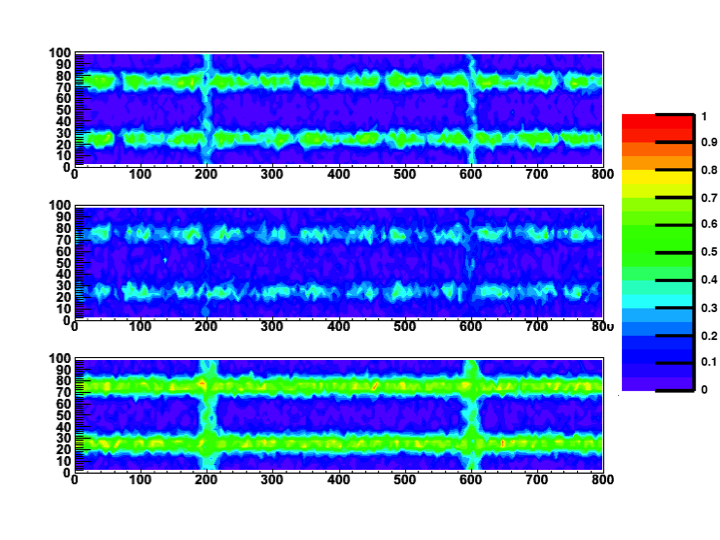}
\caption{Two-dimension probability of charge sharing between two neighboring cells for the 
FBK (top), STA (middle) and PLA (bottom) sensors, from the Eudet data at normal incidence.}
\label{fig:2dchargesharing}
\end{figure}

\subsection{Overall charge sharing probability versus tilt angle}

\begin{figure}[h]
\centering
\includegraphics[scale=0.4]{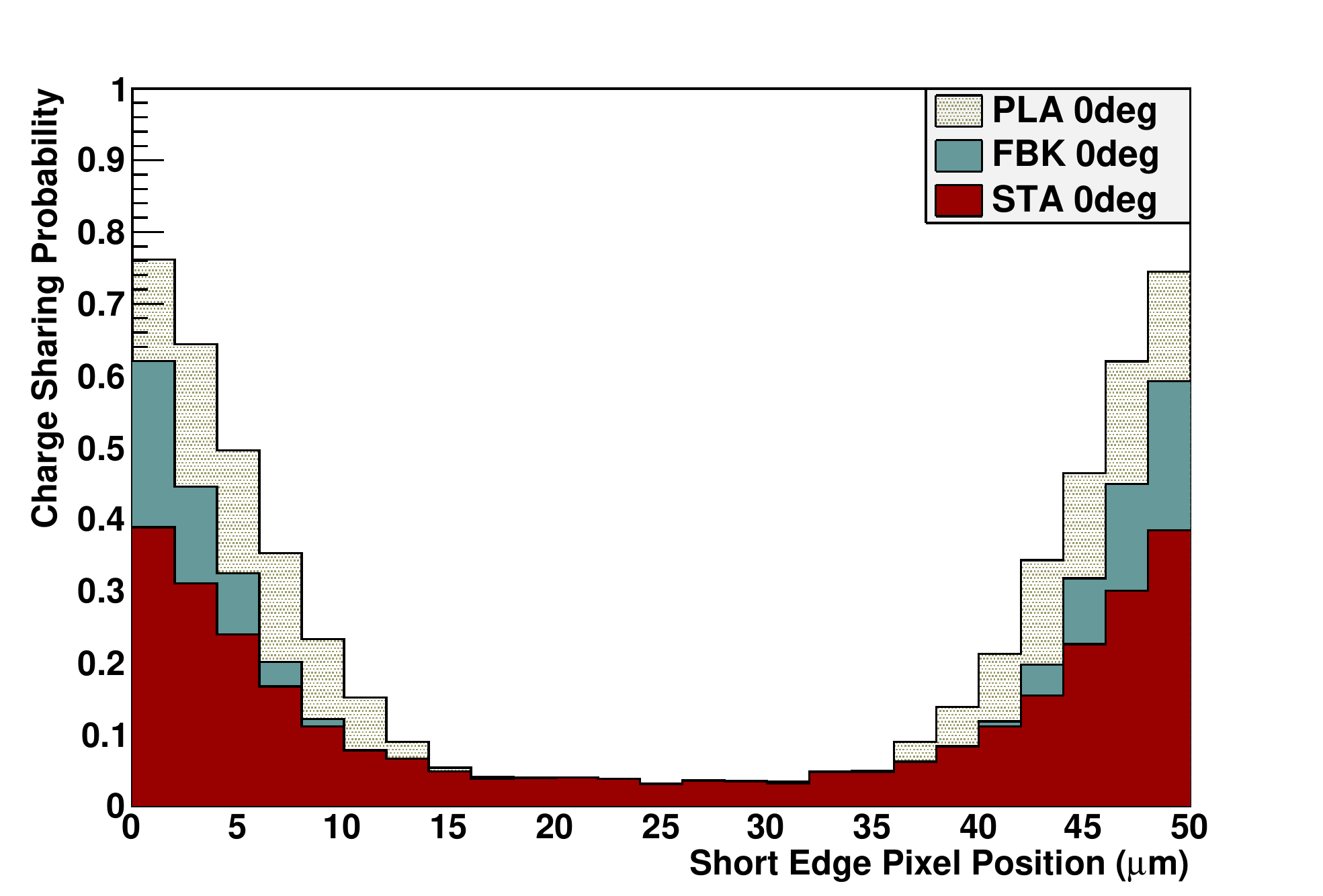}
\caption{Charge sharing probability as a function of the hit position in the $50 \mu m$ 
direction of the pixel cell, from the Eudet data at normal incidence.}
\label{fig:1dchargesharingprob}

\includegraphics[scale=0.4]{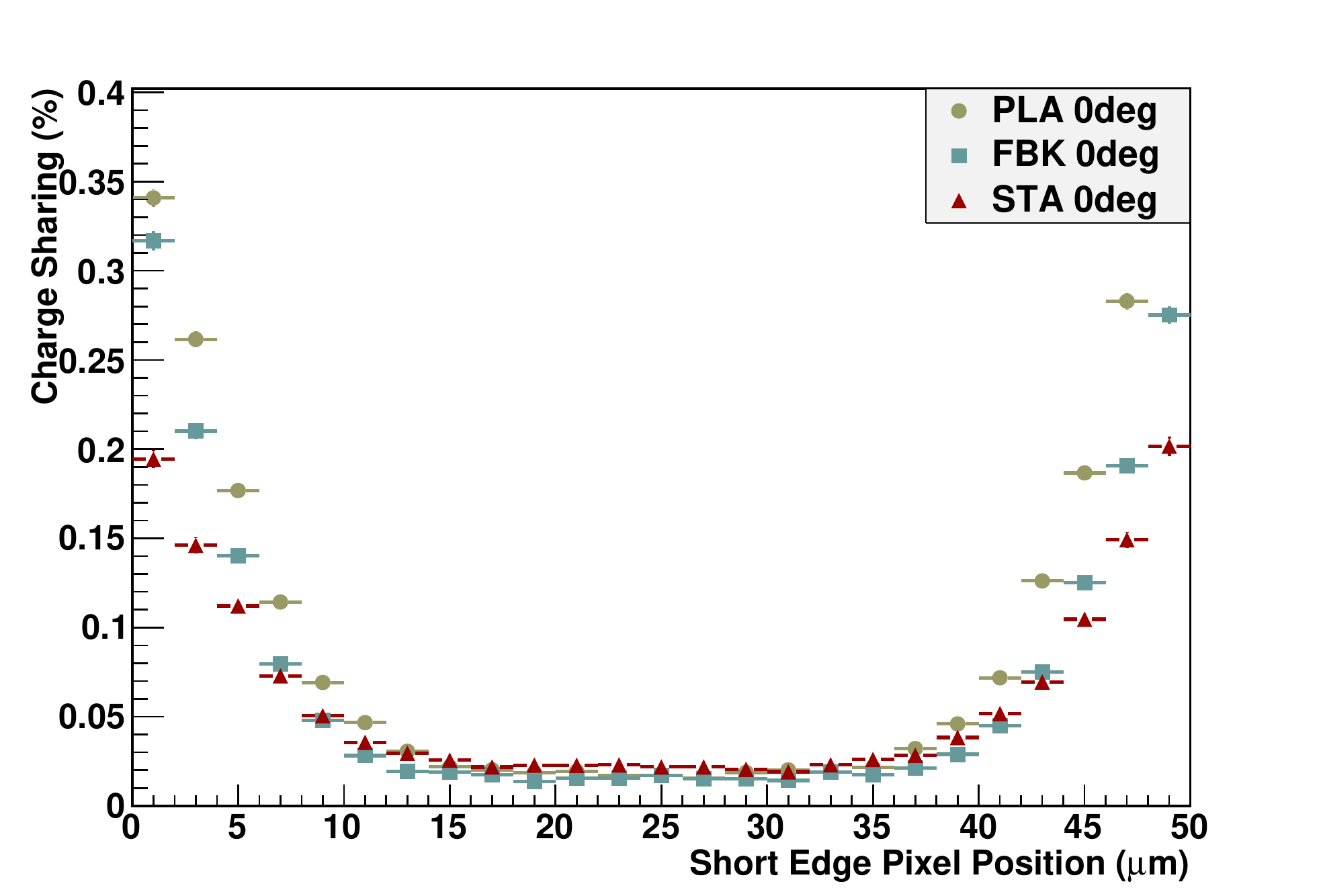}
\caption{Fraction of shared charge as a function of the hit position in the $50 \mu m$ 
direction of the pixel cell, from the Eudet data at normal incidence.}
\label{fig:1dchargesharingtot}
\end{figure}

The overall charge sharing probability, defined as the ratio of the number of tracks with 
more than one hit over the total number of tracks $N_{tracks}(>1 hit) / N_{tracks}(all)$, 
has been determined as a function of the beam incident angle for both magnetic field off 
and on. Systematic errors have been estimated, in a similar way as for the efficiencies, to be 
of the order of $3 \%$ absolute. Results are presented on Fig. \ref{fig:chargesharingvstilt}.

In the absence of magnetic field charge sharing is minimal at zero degree and has an 
expected symmetric shape versus tilt angle. Overall, charge sharing is always larger for 
the planar sensor compared to the 3D devices which have a similar behavior. Charge sharing 
is close to $100 \%$ for absolute tilt angles larger than 20 degrees. When subject to a 
magnetic field, charge sharing for the planar sensor is minimum at a value corresponding 
to the Lorentz angle. Our fitted value $(-7.4 \pm 0.4)^o$ is in excellent agreement with the ATLAS 
measurement $(-7.6 \pm 0.6)^o$ \cite{atlaslorentz} when taking into account B-field and 
temperature corrections.

We will note again the negligible effect of the magnetic fielf on the 3D sensors which shows similar 
behavior with field off and on.

\subsection{Charge sharing between neighboring cells}

Charge sharing between neighboring cells is illustrated in Fig. \ref{fig:2dchargesharing}, 
\ref{fig:1dchargesharingprob} and \ref{fig:1dchargesharingtot}. Fig. \ref{fig:2dchargesharing} 
shows the two-dimensional probability of charge sharing over two neighboring cells from the Eudet 
data at normal incidence. Fig. \ref{fig:1dchargesharingprob} shows the projection on the 
$50 \mu m$ direction of the pixel cell. As expected, charge sharing occurs predominantly in a 
narrow region separating two pixel cells and is clearly larger for the planar sensor to the 3D 
sensors. The full-3D sensor has less charge sharing compared to the FBK sensor as there is no electric 
field component perpendicular to the magnetic field. Charge sharing probability values do not reach 
100$\%$  at the edges of the cells due to tracking resolution. Additional information is found on 
Fig. \ref{fig:1dchargesharingtot} which shows the fraction of the charge shared between two 
neighboring cells as a function of the hit position in the short $50 \mu m$ direction.

%% file: conclusion.tex
\section{Conclusion}
\label{sec:conclusion}

Full and partial 3D pixel detectors have been tested, with and without a 1.6T magnetic field, in high 
energy pion beams at the CERN SPS North Area in 2009. Sensors characteristics have been measured as a 
function of the beam incident angle and compared to a regular planar pixel device. Overall full and 
partial 3D devices have similar behavior. Magnetic field has no sizeable effect on 3D performances. 
Due to electrode inefficiency 3D devices exhibit some loss of tracking efficiency for normal incident 
tracks but recover full efficiency with tilted tracks. As expected due to the electric field configuration 
3D sensors have little charge sharing between cells.

%% file: 3D-sensors-tb2009.bbl
\begin{thebibliography}{00}

\bibitem{ibl}
T.~Flick, Proceedings of Science (VERTEX 2009), Paper 033.

\bibitem{atlas3dcoll}
C. Da Vi{\`a} {\it et al.}, http://atlas-highlumi-3dsensor.web.cern.ch/atlas-highlumi-3dsensor/.

\bibitem{ref3dfull}
S. Parker, C. Kenney and J. Segal, Nucl. Instr. and Meth. A{\bf 395}, 328 (1997). 
C. Da Vi{\`a} {\it et al.}, IEEE Trans. Nucl. Sci. 56 (2009) 505. 

\bibitem{irstfbk}
G.F. Dalla Betta, et al., Nucl. Instr. and  Meth. A, doi:10.1016/j.nima.2010.04.079.


\bibitem{pellenim}
P. Hansson {\it et al}, Nucl. Instr. and Meth. A (2010), doi:10.1016/j.nima.2010.06.321.

\bibitem{bat}
J. Treus {\it et al}, Nucl. Instr. and Meth.\  A{\bf 490}, 112 (2002).

\bibitem{kyrrethesis}
K. Sj{\o}b{\ae}k, thesis (2010), Oslo University. \newline  http://www.duo.uio.no/sok/work.html?WORKID=105255.

\bibitem{morpurgo}
M.~Morpurgo, Cryogenics, 411-414 (Jul 2009).

\bibitem{eudet}
D.~Haas, Proc. of the LCWS2007, (2007),{\sc http://www.eudet.org}.

\bibitem{mimosa}
J. Baudot {\it et al.}, NSS Conference Record, 2009 IEEE, 1169. 10.1109/NSSMIC.2009.5402399.

\bibitem{Kalman}
R. ~Fr\"uwirth, Nucl. Instr. and Meth.\ A{\bf 262}, 444 (1987).

\bibitem{CKF}
R. ~Mankel, Nucl. Instr. and Meth. A{\bf 395}, 169 (1997).

\bibitem{fei3}
I.~Peric {\it et al.}, Nucl. Instr. and Meth.  A {\bf 565}, 178 (2006).

\bibitem{daf}
R. Fr\"uhwirth and A. Strandlie, Computer Physics Communications, Volume 120, Issues 2-3, August 1999, Pages 197-214.

\bibitem{stananofab}
C. Da Vi{\`a} {\it et al.}, Nucl. Instr. and Meth. A 604 (2009) 505.

\bibitem{atlaspixel}
G.~Aad~(ATLAS Collaboration), JINST3, P07007, {\bf 7} (2008).

\bibitem{mathes}
M. Mathes {\it et al.}, IEEE Trans. Nucl. Sci. 55 (2008) 3731.


\bibitem{elecfill}
C. Kenney, Private communication.

\bibitem{atlaslorentz}
P. Behera {\it et al}, ATL-COM-INDET-2010-041.


\end{thebibliography}
